\documentclass[12pt]{article}
\usepackage{amsfonts,graphicx,amssymb,amsmath,amsthm,epsfig}
\usepackage{float}
\allowdisplaybreaks
\begin{document}
\title{\bf Pilgrim and Generalized Ghost Pilgrim Dark Energy Models
in Modified Symmetric Teleparallel Gravity}
\author{M. Sharif \thanks {msharif.math@pu.edu.pk}~and
Madiha Ajmal \thanks {madihaajmal222@gmail.com} \\
Department of Mathematics and Statistics, The University of Lahore,\\
1-KM Defence Road Lahore-54000, Pakistan.}

\date{}
\maketitle
\begin{abstract}
In this paper, we investigate $f(Q)$ gravity ($Q$ represents the
non-metricity) to explore its cosmological implications of the two
non-interacting dark energy models. We use the correspondence
scenario to find the pilgrim and generalized ghost pilgrim dark
energy $f(Q)$ gravity models for a Friedmann-Robertson-Walker
universe with pressureless matter and a power-law scale factor. The
reconstructed models explain the observed rapid expansion of the
universe using astronomical observations. We also examine the
physical properties of the model, including its equation of state
parameter, $(\omega_{D}-\omega^{\prime}_{D})$ and $(r-s)$-planes and
the squared speed of sound. It is found that our results are
consistent with the interacting pilgrim dark energy model in the
same gravity.
\end{abstract}
\textbf{Keywords}: Cosmological evolution, Dark
energy, $f(Q)$ gravity.\\
\textbf{PACS}: 64.30.+t; 95.36+x; 04.50.Kd.\\

\section{Introduction}

Numerous characteristics of the universe have been described by the
general theory of relativity (GR) using a variety of observational
data. Recent astrophysical findings such as Supernovae type-Ia
\cite{5}, large-scale structure \cite{6} combined with the baryon
acoustic oscillations \cite{8} and cosmic microwave background
radiations \cite{9} indicate that the universe is currently
experiencing accelerated expansion. There is a strong evidence
suggesting that this universe is mainly defined by the mysterious
components, i.e., dark matter (DM) and dark energy (DE). In GR,
adding a cosmological constant into the field equations helps us to
better understand the mysterious properties of DE. However, this
constant introduces challenges like the fine-tuning and coincidence
problems \cite{9a}. Several modified gravity theories (MGTs) have
been suggested to address the complexities in GR. There are two
established approaches to define the cosmic acceleration. One
significant aspect is the introduction of DE with large negative
pressure in GR \cite{6} and the other is the extension and
modification of the GR action. In recent years, experiments have
shown that MGTs effectively describe early phenomena such as
inflation and late-period acceleration.

In order to investigate the characteristics of DE, various efforts
have been made such as modification of gravity and DE models.
Dynamical DE models are attained by altering the matter component,
incorporating entities such as phantom, quintessence, Chaplygin gas,
among others. Various models have been developed based on energy
densities to comprehend various stages of the cosmic evolution.
According to recent research, this can be explained by adding a new
degree of freedom or by assuming a new parameter. The new DE model
may have numerous unfamiliar characteristics and can give rise to
fresh challenges in existing literature. Our initial approach is to
address the DE problem without bringing in additional degrees of
freedom beyond those already recognized. A new category of models
called ghost DE (GDE) has attracted considerable interest. According
to this theory, the Veneziano ghost field is believed to be
accountable for the recent expansion of the universe \cite{10}. The
energy density of the GDE model is expressed as $\rho_D=\alpha H$,
where the constant $\alpha$ has dimensions of $[energy]^{3}$. Cai et
al. \cite{11} concluded that incorporating the second-order term in
the GDE model is crucial for precisely depicting the dynamics of the
early universe. This modification characterizes the model for GGDE
as $\rho_D=\alpha H+\beta H^{2}$, where $\beta$ represents another
constant with dimensions of $[energy]^{2}$.

Wei \cite{12} proposed the pilgrim DE (PDE), emphasizing a
phantom-like universe to avoid the formation of black holes. The PDE
model is given as
\begin{equation}\label{1}
\rho_D=\alpha (H)^{\psi}, \quad\psi\leq 2,
\end{equation}
$\psi$ is a dimensionless constant. The MGTs have also been used to
explore the cosmic implications of the GGPDE model. The generalized
ghost PDE (GGPDE) model is the commonly accepted PDE extension of
GGDE. The energy density of the GGPDE model can be described by the
following equation
\begin{equation}\label{2}
\rho_D=(\alpha H+\beta H^{2})^{\psi}.
\end{equation}

Sharif and Jawad \cite{13} explored the proposal of PDE in an
interacting framework with CDM using three cutoffs. Sharif and
Zubair \cite{15} formulated a PDE model in $f(R)$ gravity using
various IR cut-offs and reconstructed $f(R)$ model to demonstrate
its unique behavior in future cosmic evolution. Sharif and Rani
\cite{16} investigated the mysterious nature of PDE in $f(T)$
gravity ($T$ represents the torsion scalar) using Hubble horizon as
an IR-cutoff and investigated phase planes as well as cosmological
parameters. Chattopadhyay et al. \cite{17} studied the puzzling
characteristics of PDE in $f(T,T_{G})$ gravity ($T_{G}$ denotes the
teleparallel equivalent of the Gauss-Bonnet term) and found an
extremely active phantom-like behavior, which is essential for
preventing black hole formation. Sharif and Nazir \cite{21}
investigated GGPDE to examine black hole creation in the context of
$f(T,T_{G})$ gravity. Sharif and Nawazish \cite{21a} studied various
standard cosmological parameters and stability criteria in GGPDE
$f(R)$ gravity model. Sharif and Saba \cite{21b} analyzed the
cosmological behavior of the reconstructed PDE and GGPDE models
using cosmic diagnostic parameters, phase planes and the squared
speed of sound ($\nu_{s}^{2}$) in $f(G,\mathcal{T})$ gravity
($\mathcal{T}$ is trace of the energy-momentum tensor (EMT)).

General relativity is described through Riemannian geometry which
requires the affine connection of spacetime manifold to adhere the
metric compatibility known as the Levi-Civita connection \cite{22}.
A manifold can support different options for affine connection and
different connections could result in various representations of
gravity. As a result, they may provide varying viewpoints and
understanding of the phenomenon. The GR specifies that the
Levi-Civita connection requires both the non-metricity $Q$ and
torsion $T$ to be zero, while keeping the curvature as a fundamental
geometrical object. By relaxing these constraints, it is possible to
develop the theories of gravity rooted in non-Riemannian geometry,
where curvature, torsion and non-metricity can all be non-zero. The
teleparallel equivalent of GR (TEGR) \cite{25,26} can be stated by
selecting a connection that allows non-zero torsion but does not
require curvature or non-metricity. By considering a flat spacetime
without torsion but with non-zero non-metricity, one could develop
the symmetric teleparallel formulation of GR (STGR)
\cite{29}-\cite{34}.

Researchers are much motivated in exploring the non-Riemannian
geometry, for example, $f(Q)$ theory. The motivation behind this
theory is to examine its theoretical significance, consistency with
observed data and importance in the study of cosmic phenomena. This
theory examines theoretical implications derived from cosmic domains
and empirical observations. Lazkoz et al. \cite{43} examined the
limitations of $f(Q)$ gravity through the utilization of polynomial
expressions in relation to red-shift. They also studied the energy
conditions for two different models in this gravity. Bajardi et al.
\cite{45} derived the cosmic wave function using Hamiltonian
formalism in this framework. Shekh \cite{46} conducted a dynamic
analysis of the holographic DE model within the same theoretical
framework. Frusciante \cite{47} presented a specific model in this
gravity which exhibited similarities to the $\Lambda$CDM model on a
foundational level. Lymperis \cite{49} investigated the cosmic
evolution in the presence and absence of cosmological constant
$\Lambda$ with phantom DE. Dimakis et al. \cite{50} studied cosmic
evolution with phantom DE both in the presence and absence of the
cosmological constant. Khyllep et al. \cite{51} examined the
universe accelerated expansion using power-law and exponential
models of $f(Q)$ theory. In a recent paper \cite{52}, we have
explored the cosmography of GGDE in the same gravity. Sharif et al.
\cite{53} explored the idea of a cosmological bounce in
non-Riemannian geometry.

This paper explores the correspondence scheme involving PDE and
GGPDE models by using reconstruction technique within $f(Q)$
gravity. The structure of the paper is as follows. Section
\textbf{2} discusses the FRW universe with fluid sources
non-interacting between DM and DE and the field equation of $f(Q)$
gravity. Sections \textbf{3} and \textbf{4} provide a discussion on
cosmographic observations using cosmic diagnostic parameters and
phase planes for the reconstructed PDE and GGPDE $f(Q)$ gravity
models, respectively. Finally, we present our conclusions in section
\textbf{5}.

\section{FRW Universe Model and $f(Q)$ Gravity}

The line element that characterizes a spatially homogeneous and
isotropic model of the universe is expressed as
\begin{equation}\label{3}
ds^{2}=-dt^{2}+a^{2}(t)(d\mathbf{x}^{2}+d\mathbf{y}^{2}+d\mathbf{z}^{2}),
\end{equation}
where a scale factor is denoted by $a(t)$. The total EMT for DE and
DM is
\begin{equation}\label{4}
\hat{\mathcal{T}_{\psi\gamma}}=\mathcal{T}_{\psi\gamma}+\tilde{\mathcal{T}}_{\;\psi\gamma},
\end{equation}
where the EMTs for pressureless DM and DE are denoted by
$\mathcal{T}_{\;\psi\gamma}$ and
$\tilde{\mathcal{T}}_{\;\psi\gamma}$, defined as
$\mathcal{T}_{\;\psi\gamma}=(\rho_{\mathbf{m}})u_{\psi}u_{\gamma}$
and
$\tilde{\mathcal{T}}_{\;\psi\gamma}=(\rho_{D}+p_{D})u_{\psi}u_{\gamma}+p_{D}g_{\psi\gamma}$,
respectively, $u_{\gamma}$ is the four velocity, $p_{D}$ denotes the
pressure of DE, and $\rho_{\mathbf{m}}$ and $\rho_{D}$ indicate the
energy densities of DM and DE, respectively. The following
expressions denote the energy densities in fractional form for DE
and DM as
\begin{equation}\label{5}
\Omega_{\mathbf{m}}=\frac{\rho_{\mathbf{m}}}{\rho_{cr}}=
\frac{\rho_{\mathbf{m}}}{3H^{2}}, \quad
\Omega_{D}=\frac{\rho_{D}}{\rho_{cr}}=\frac{\rho_{D}}{3H^{2}},
\end{equation}
implying that $1$ can be expressed as the sum of $\Omega_{D}$ and
$\Omega_{\mathbf{m}}$, $\rho_{cr}$ is the critical density. The
continuity equations for the non-interacting DM and DE are
\begin{eqnarray}\label{6}
\dot{\rho}_{\mathbf{m}}+3H(\rho_{\mathbf{m}})&=&0,\\\label{7}
\dot{\rho}_{D}+3H(\rho_{D}+p_{D})&=&0.
\end{eqnarray}
We assume the scale factor in power-law form as
\begin{equation}\label{8}
a(t)=a_{0}t^{m},
\end{equation}
where $m$ and $a_0$ are arbitrary constants and $a_0$ has a current
value of $1$. The power-law form simplifies the differential
equations governing cosmological dynamics, making analytical
solutions more accessible. This aids in understanding the
qualitative behavior of the universe's expansion and the evolution
of DE. When $ m > 1$, the scale factor indicates accelerated
expansion, aligning with current observations of the universe's
acceleration. By examining how $m$ influences the universe's
expansion and comparing it with observational data, these models can
offer deeper insights into the nature of DE and the fundamental
properties of gravity. The parameter $m$ is crucial for determining
how quickly the universe transitions between different phases of
expansion, such as from decelerated to accelerated expansion. This
is crucial for matching the model to different cosmological epochs,
including the matter-dominated era and the dark energy-dominated
era. Fitting the power-law scale factor to cosmological data allows
researchers to test the compatibility of the PDE and GGPDE $f(Q)$
models with observations. This helps in validating or constraining
these models.

The values of $H$, its derivative and the non-metricity in terms of
this scale factor are given as follows
\begin{equation}\label{9}
H=\frac{\dot{a}}{a}=\frac{m}{t}, \quad\dot{H}=-\frac{m}{t^{2}},
\quad Q=6H^{2}=6 \frac{m^2}{t^2},
\end{equation}
where, dot represents derivative with respect to $t$. Integrating
Eq.\eqref{6}, we have
\begin{equation}\label{10}
\rho_{\mathbf{m}}=\xi(a)^{-3}=\xi(t^m)^{-3},
\end{equation}
where $\xi$ is an integration constant.

The action for $f(Q)$ gravity is expressed as \cite{32}
\begin{equation}\label{11}
S=\int\bigg(\frac{1}{2k}f(Q)+\mathcal{L}_{\mathbf{m}}\bigg)
\sqrt{-g}d^{4}x,
\end{equation}
where $g$ represents determinant of the metric tensor,
$\mathcal{L}_{\mathbf{m}}$ stands for the matter Lagrangian density
and $Q$ is described as
\begin{equation}\label{12}
Q=-g^{\gamma\psi}(\mathbb{L}^{\mu}_{~\nu\gamma}\mathbb{L}^{\nu}_{~\psi\mu}
-\mathbb{L}^{\mu}_{~\nu\mu}\mathbb{L}^{\nu}_{~\gamma\psi}).
\end{equation}
The Levi-Civita connection in symmetric connections can be written
using the deformation tensor as
$\Gamma^{\mu}_{\nu\varsigma}=-\mathbb{L}^{\mu}_{\;\nu\varsigma}$,
where
\begin{equation}\label{13}
\mathbb{L}^{\mu}_{\;\nu\varsigma}=-\frac{1}{2}g^{\mu\lambda}
(\nabla_{\varsigma}g_{\nu\lambda}+\nabla_{\nu}g_{\lambda\varsigma}
-\nabla_{\lambda}g_{\nu\varsigma}).
\end{equation}
The traces of the $Q$ tensor are defined as
\begin{equation}\label{14}
Q_{\mu}=Q^{~\psi}_{\mu~\psi},\quad
\tilde{Q}_{\mu}=Q^{\psi}_{~\mu\psi}.
\end{equation}
The superpotential can be written as
\begin{equation}\label{15}
\mathbb{P}^{\mu\psi\gamma}=\frac{1}{4}\big[-Q^{\mu\psi\gamma}+Q^{\psi\mu\gamma}
+Q^{\gamma\mu\psi}+Q^{\psi\mu\gamma}-\tilde{Q}_{\mu}g^{\psi\gamma}
+Q^{\mu}g^{\psi\gamma}\big].
\end{equation}
Consequently, the relation for $Q$ becomes \cite{32a}
\begin{equation}\label{16}
Q=-Q_{\mu\gamma\psi}\mathbb{P}^{\mu\gamma\psi}=-\frac{1}{4}(-Q^{\mu\psi\rho}Q_{\mu\psi\rho}
+2Q^{\mu\psi\rho}Q_{\rho\mu\psi}-2Q^{\rho}\tilde{Q}_{\rho}+Q^{\rho}Q_{\rho}).
\end{equation}
The corresponding field equations of $f(Q)$ gravity take the form
\begin{equation}\label{17}
\frac{-2}{\sqrt{-g}}\nabla_{\gamma}(f_{Q}\sqrt{-g}
P^{\mu}_{~\gamma\psi})-\frac{1}{2}f g_{\gamma\psi}-f_{Q}
(P_{\gamma\mu\nu}Q_{\psi}^{~\mu\nu}-2Q^{\mu\nu}_{~~~\gamma}
P_{\mu\nu\psi})=k^{2} T_{\gamma\psi},
\end{equation}
where $f_{Q}=\frac{\partial f(Q)}{\partial Q}$. The modified
Friedmann equations for $f(Q)$ gravity are
\begin{equation}\label{18}
2\dot{H}+3H^{2}=p_D +p_{\mathbf{m}},\quad 3H^{2}=\rho_D
+\rho_{\mathbf{m}},
\end{equation}
where
\begin{eqnarray}\label{19}
\rho_D&=&-6H^{2}f_{Q}+\frac{f}{2},\\\label{20}
p_D&=&2f_{Q}\dot{H}+2Hf_{QQ}+6H^{2}f_{Q}-\frac{f}{2}.
\end{eqnarray}

\section{Reconstruction of PDE $f(Q)$ Model}

In this section, we use a method to reconstruct the PDE $f(Q)$ model
by equating the corresponding densities. Using Eqs.\eqref{1} and
\eqref{19}, we have
\begin{equation}\label{21}
\frac{f}{2}-6H^{2}f_{Q}=\alpha H^{\psi}.
\end{equation}
The PDE model often involves modifications to standard gravitational
theories to accommodate its effects on cosmic dynamics. This
equation represents the first-order linear differential equation in
$Q$ and its solution is obtained as
\begin{equation}\label{22}
f(Q)=\frac{6^{-\frac{\psi }{2}} \bigg(c \sqrt{Q} 6^{\psi /2} (\psi
-1)-2 \alpha  Q^{\psi /2}\bigg)}{\psi -1}.
\end{equation}
In terms of $t$, this model can be obtained by substituting
Eq.\eqref{9} in \eqref{22} as
\begin{figure}
\epsfig{file=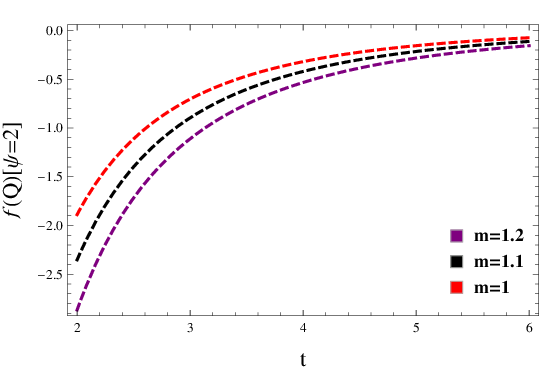,width=0.5\linewidth}
\epsfig{file=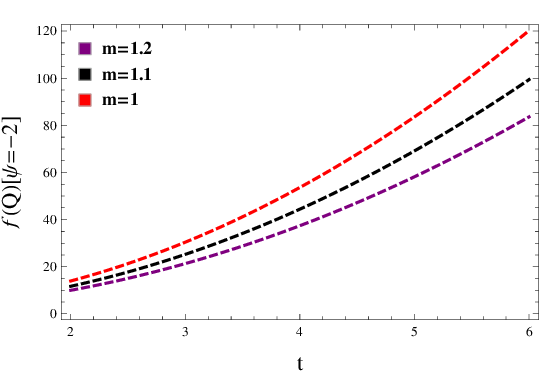,width=0.5\linewidth}\caption{Graphs show the
relationship between $f(Q)$ ($\psi=\pm2$) and $t$.}
\end{figure}
\begin{equation}\label{23}
f(Q)=\frac{\sqrt{6} c (\psi -1) \sqrt{\frac{m^2}{t^2}}-2 \alpha
\bigg(\frac{m^2}{t^2}\bigg)^{\psi /2}}{\psi -1}.
\end{equation}
Figure \textbf{1} shows the behavior of the reconstructed PDE $f(Q)$
gravity model for three various values $m=1.2,~ 1.1,~ 1$, indicating
decrease with increasing values of $t$ for $\psi=2$. On the other
hand, for $\psi=-2$, $f(Q)$ increases with increasing values of $t$.
The values of $\rho_D$ and $p_D$ are found by substituting
Eq.\eqref{22} in \eqref{19} and \eqref{20} as
\begin{eqnarray}\nonumber
\rho_D&=&\alpha  6^{-\frac{\psi }{2}} Q^{{\psi }{2}}+c,\\\nonumber
p_D&=&\frac{1}{Q^2 (\psi -1)}\bigg[2^{-\frac{\psi}{2}-1}
3^{-\frac{\psi}{2}} \bigg(c Q^{\frac{3}{2}} 2^{\frac{\psi}{2}+1}
3^{\frac{\psi}{2}} (\psi -1) \big(\dot{H}+3 H^2\big)
\\\nonumber&-&c H\sqrt{Q}
6^{\frac{\psi}{2}}(\psi -1)+c Q^{\frac{5}{2}} 6^{\frac{\psi}{2}}
(\psi -1)-4 \alpha  \psi \big(\dot{H}+3 H^2\big) Q^{\frac{\psi
}{2}+1} \\\nonumber&-&2 \alpha H (\psi -2) \psi Q^{\frac{\psi}{2}}-2
\alpha Q^{\frac{\psi }{2}+2}\bigg)\bigg].
\end{eqnarray}
Inserting Eq.\eqref{9} into the above equations, we can represent
them in the form of a power-law as
\begin{eqnarray}\nonumber
\rho_D&=&\alpha\bigg(\frac{m^2}{t^2}\bigg)^{\psi /2}+c,
\\\nonumber
p_D&=&-\frac{1}{72 m^3 (\psi -1)}\bigg[-12 m^2 \bigg(2 \alpha  \psi
\bigg(\frac{m^2}{t^2}\bigg)^{\psi /2}-\sqrt{6} c (\psi -1)
\sqrt{\frac{m^2}{t^2}}\bigg)\\\nonumber&+&t^3 \bigg(\sqrt{6} c (\psi
-1) \sqrt{\frac{m^2}{t^2}}+2 \alpha  (\psi -2) \psi
\bigg(\frac{m^2}{t^2}\bigg)^{\psi /2}\bigg)
\\\nonumber&+&\bigg(\alpha (\psi +1)\bigg(\frac{m^2}{t^2}\bigg)^{\psi /2}-\sqrt{6} c
(\psi -1) \sqrt{\frac{m^2}{t^2}}\bigg)72 m^3 \bigg].
\end{eqnarray}
Figures \textbf{2} and \textbf{3} show the behavior of $\rho_D$ and
$p_D$ over time. The energy density $\rho_D$ exhibits a positive
trend, while $p_D$ illustrates a negative pattern for all values of
$m$ and $\psi$, consistent with the behavior expected for DE.
\begin{figure}
\epsfig{file=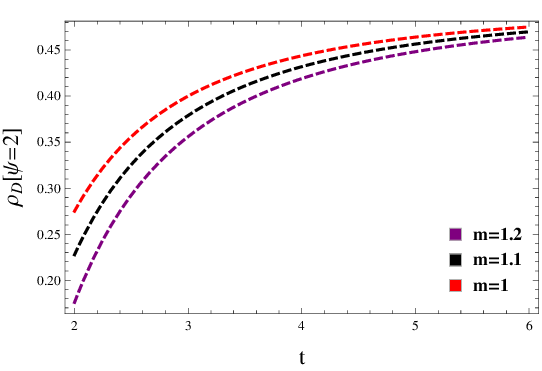,width=.5\linewidth}
\epsfig{file=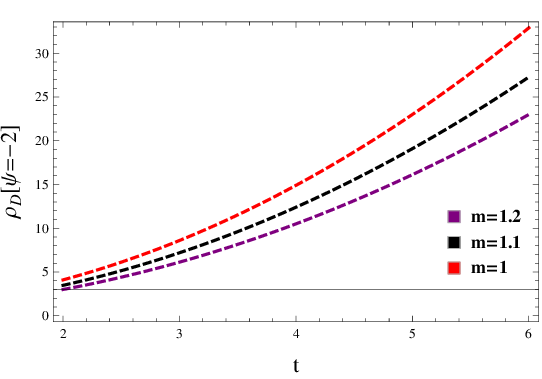,width=.49\linewidth}\caption{Graphs of $\rho_D$
($\psi=\pm2$) against $t$.}
\end{figure}
\begin{figure}
\epsfig{file=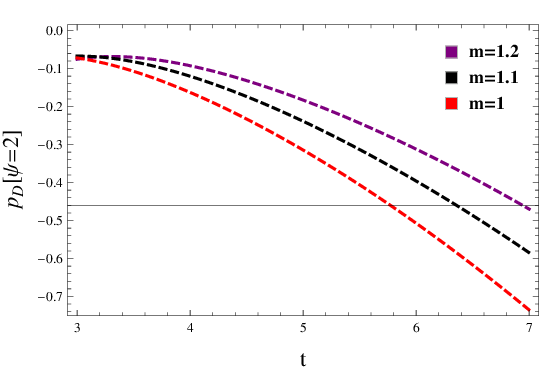,width=.49\linewidth}
\epsfig{file=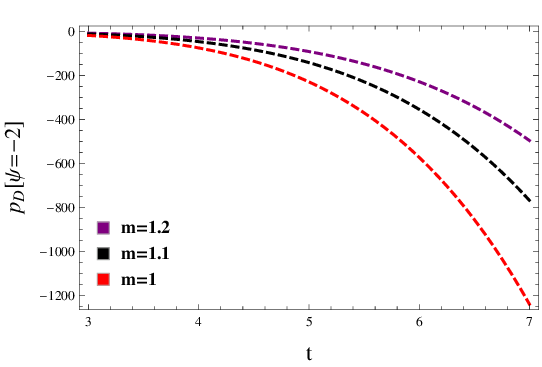,width=.5\linewidth}\caption{Graphs of $p_D$
($\psi=\pm2$) versus $t$.}
\end{figure}

Now, we illustrate the behavior of different cosmological
parameters, such as the equation of state parameter (EoS) parameter
and phase planes. We also analyze the stability of this model. The
EoS is defined by the ratio ($\omega_{D}=\frac{p_{D}}{\rho_{D}}$).
For different forms of matter and energy, $\omega_{D}$ takes on
different values. For matter-dominated regions, i.e.,
non-relativistic matter (dust) $(\omega_{D}=0)$, radiation
$(\omega_{D}=\frac{1}{3})$ and stiff matter $(\omega_{D}=1)$. For DE
models, the EoS parameter helps in understanding the nature of DE,
whether it behaves like a vacuum, quintessence and phantom energy.
For instance, $\omega_{D}=-1$ corresponds to a vacuum, while
$\omega_{D}<-1$ indicates phantom energy,
$-1<\omega_{D}<-\frac{1}{3}$ leads to quintessence phase. Thus we
have
\begin{eqnarray}\nonumber
\omega_{D}&=&-\frac{1}{72 m^3 (\psi -1) \bigg(\alpha t^{3 m}
\bigg(\frac{m^2}{t^2}\bigg)^{\psi /2}+\xi \bigg)}\bigg[t^{3 m}
\bigg(-12 m^2 \bigg(2 \alpha  \psi \bigg(\frac{m^2}{t^2}\bigg)^{\psi
/2}\\\nonumber&-&\sqrt{6} c (\psi -1)
\sqrt{\frac{m^2}{t^2}}\bigg)+t^3 \bigg(\sqrt{6} c (\psi -1)
\sqrt{\frac{m^2}{t^2}}+2 \alpha  (\psi -2) \psi
\bigg(\frac{m^2}{t^2}\bigg)^{\frac{\psi}{2}}\bigg)\\\nonumber&+&72
m^3 \bigg(\alpha (\psi +1) \bigg(\frac{m^2}{t^2}\bigg)^{\psi
/2}-\sqrt{6} c (\psi -1) \sqrt{\frac{m^2}{t^2}}\bigg)\bigg)\bigg].
\end{eqnarray}
Figure \textbf{4} shows the EoS parameter for three distinct values
of $m$, indicating $\omega_{D}<-1$. This suggests that the model
incorporates phantom field DE for both $\psi=2$ and $\psi=-2$, and
fulfills the PDE phenomenon.
\begin{figure}
\epsfig{file=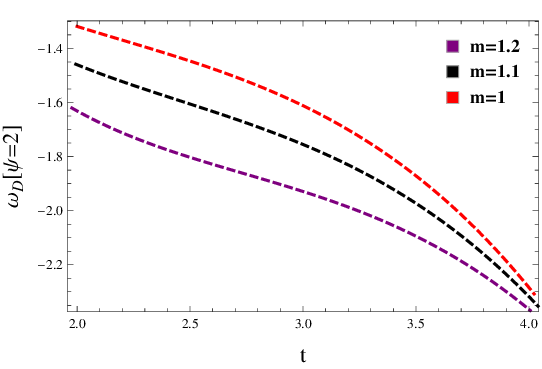,width=.5\linewidth}
\epsfig{file=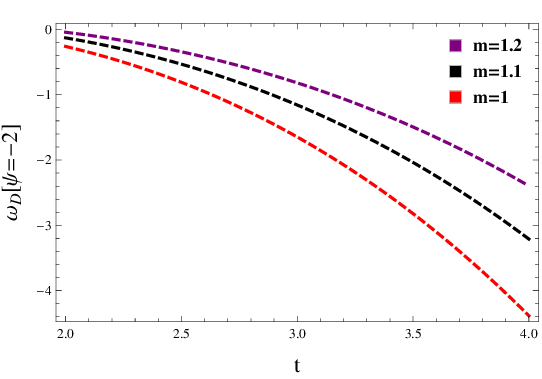,width=.49\linewidth}\caption{Plots of
$\omega_{D}$ ($\psi=\pm2$) against $t$.}
\end{figure}

The ($\omega_{D}-\omega^{\prime}_{D}$)-plane \cite{55} (prime means
derivative with respect to $Q$) categorizes different DE scenarios
into distinct regions such as thawing
($\omega_{D}<0,~\omega^{\prime}_{D}>0$) and freezing
($\omega_{D}<0,~\omega^{\prime}_{D}<0$), based on their evolutionary
trajectories. Here we have
\begin{eqnarray}\nonumber
\omega^{\prime}_{D}&=&\frac{1}{432 \sqrt{6} m^5 (\psi -1)
\bigg(\alpha t^{3 m} \bigg(\frac{m^2}{t^2}\bigg)^{\psi /2}+\xi
\bigg)^2}\bigg[t^{3 m+2} \bigg(-12 m^2 \bigg(t^{3 m} \bigg(2
\sqrt{6} \\\nonumber&\times&\alpha ^2 \psi
\bigg(\frac{m^2}{t^2}\bigg)^{\psi }-3 \alpha  c \big(\psi ^2-1\big)
\bigg(\frac{m^2}{t^2}\bigg)^{\frac{\psi +1}{2}}\bigg)-\xi \bigg(3 c
(\psi -1) \sqrt{\frac{m^2}{t^2}}+\sqrt{6}\\\nonumber&\times& \alpha
(\psi -2) \psi
\bigg(\frac{m^2}{t^2}\bigg)^{\frac{\psi}{2}}\bigg)\bigg)+t^3
\bigg(\alpha t^{3 m} \bigg(\frac{m^2}{t^2}\bigg)^{\frac{\psi}{2}}
\bigg(3 c \big(\psi ^2+2 \psi -3\big)
\sqrt{\frac{m^2}{t^2}}\\\nonumber&+&4 \sqrt{6} \alpha  (\psi -2)
\psi  \bigg(\frac{m^2}{t^2}\bigg)^{\frac{\psi}{2}}\bigg)-\xi
\bigg(\sqrt{6} \alpha  \psi  \bigg(\psi ^2-6 \psi +8\bigg)
\bigg(\frac{m^2}{t^2}\bigg)^{\frac{\psi}{2}}-9 c
\\\nonumber&\times&(\psi -1) \sqrt{\frac{m^2}{t^2}}\bigg)\bigg)+36
\alpha  m^3 \psi \bigg(\frac{m^2}{t^2}\bigg)^{\psi /2} \bigg(2 t^{3
m} \bigg(\sqrt{6} \alpha  \bigg(\frac{m^2}{t^2}\bigg)^{\psi /2}-3 c
\\\nonumber&\times&(\psi -1) \sqrt{\frac{m^2}{t^2}}\bigg)-\sqrt{6} \xi  (\psi
-1)\bigg)\bigg)\bigg].
\end{eqnarray}
Figure \textbf{5} indicates that
$\omega_{D}<0,~\omega^{\prime}_{D}<0$ for all values of $m$ and PDE
parameters, demonstrating the existence of the freezing region. The
$(r-s)$-plane \cite{57} helps to distinguish between different
models of the universe expansion. In this plane, trajectories
falling within the range $(r<1)$ and $(s>0)$ represent epochs
dominated by phantom and quintessence DE, respectively. Conversely,
trajectories characterized by $(r>1)$ and $(s<0)$ correspond to the
Chaplygin gas model. The two dimensionless parameters for the
$(r-s)$-plane are given as
\begin{figure}
\epsfig{file=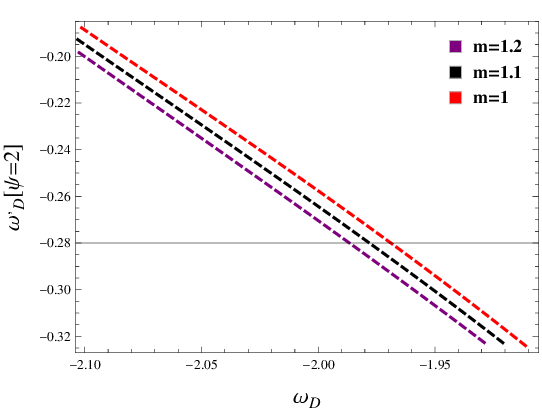,width=.45\linewidth}
\epsfig{file=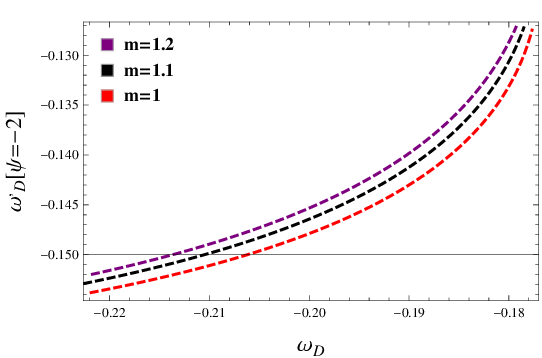,width=.515\linewidth}\caption{Graphs of
$\omega^{\prime}_{D}$ ($\psi=\pm2$) against $\omega_{D}$.}
\end{figure}
\begin{equation}\nonumber
r=\frac{\dddot{a}}{aH^{3}}, \quad s=\frac{r-1}{3(q-\frac{1}{2})}.
\end{equation}
The values of $r$ and $s$ are given in Appendix $\textbf{A}$. Figure
\textbf{6} demonstrates the Chaplygin gas model ($r>1$ and $s<0$)
for various values of $m$ and PDE parameter.
\begin{figure}\center
\epsfig{file=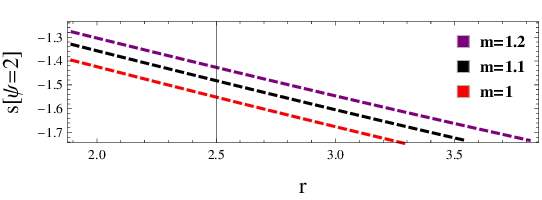,width=.5\linewidth}
\epsfig{file=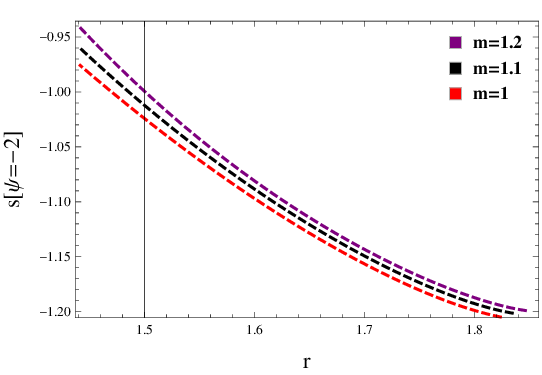,width=.45\linewidth}\caption{Plots of $r$
($\psi=\pm2$) versus $s$.}
\end{figure}

The behavior of the $\nu_{s}^{2}$ parameter is crucial in
cosmological models, dictating the stability of cosmic structures.
When the $\nu_{s}^{2}$ remains positive, this indicates stability,
while its negative value signals instability. This is given as
\begin{equation}\nonumber
\nu_{s}^{2}=\frac{\dot{p}_{D}}{\dot{\rho}_{D}}
=\frac{\rho_D}{{\dot{\rho}_{D}}}\omega^{\prime}_{D}+\omega_{D},
\end{equation}
and hence
\begin{eqnarray}\nonumber
\nu_{s}^{2}&=&-\frac{1}{5184 \alpha  m^6 (\psi -1)^2 \psi
\bigg(\alpha  t^{3 m} \bigg(\frac{m^2}{t^2}\bigg)^{\psi /2}+\xi
\bigg)^2}\bigg[t^{3 m} \bigg(\frac{m^2}{t^2}\bigg)^{-\frac{\psi
}{2}} \bigg(-12 m^2 \\\nonumber&\times&\bigg(2 \alpha  \psi
\bigg(\frac{m^2}{t^2}\bigg)^{\psi /2}-\sqrt{6} c (\psi -1)
\sqrt{\frac{m^2}{t^2}}\bigg)+t^3 \bigg(\sqrt{6} c (\psi -1)
\sqrt{\frac{m^2}{t^2}}+2 \alpha \\\nonumber&\times& (\psi -2) \psi
\bigg(\frac{m^2}{t^2}\bigg)^{\psi /2}\bigg)+\bigg(\alpha (\psi +1)
\bigg(\frac{m^2}{t^2}\bigg)^{\psi /2}-\sqrt{6} c (\psi -1)
\sqrt{\frac{m^2}{t^2}}\bigg)\\\nonumber&\times&72 m^3 \bigg)
\bigg(-12 m^2 \bigg(\xi \bigg(-\sqrt{6} c (\psi -1)
\sqrt{\frac{m^2}{t^2}}-2 \alpha  (\psi -2) \psi
\bigg(\frac{m^2}{t^2}\bigg)^{\psi /2}\bigg)\\\nonumber&+&t^{3 m}
\bigg(4 \alpha ^2 \psi \bigg(\frac{m^2}{t^2}\bigg)^{\psi }-\sqrt{6}
\alpha c \bigg(\psi ^2-1\bigg)
\bigg(\frac{m^2}{t^2}\bigg)^{\frac{\psi +1}{2}}\bigg)\bigg)+t^3
\bigg(\alpha \bigg(\frac{m^2}{t^2}\bigg)^{\psi
/2}\\\nonumber&\times& t^{3 m}\bigg(\sqrt{6} c \bigg(\psi ^2+2 \psi
-3\bigg) \sqrt{\frac{m^2}{t^2}}+8 \alpha (\psi -2) \psi
\bigg(\frac{m^2}{t^2}\bigg)^{\psi /2}\bigg)-\xi \bigg(2 \alpha \psi
\\\nonumber&\times&\bigg(\psi ^2-6 \psi +8\bigg) \bigg(\frac{m^2}{t^2}\bigg)^{\psi
/2}-3 \sqrt{6} c (\psi -1) \sqrt{\frac{m^2}{t^2}}\bigg)\bigg)+72
\alpha  m^3 \psi  t^{3 m}
\\\nonumber&\times&\bigg(\frac{m^2}{t^2}\bigg)^{\psi /2}
\bigg(\alpha  (\psi +1) \bigg(\frac{m^2}{t^2}\bigg)^{\psi
/2}-\sqrt{6} c (\psi -1) \sqrt{\frac{m^2}{t^2}}\bigg)\bigg)\bigg].
\end{eqnarray}
Figure \textbf{7} indicates that $\nu_{s}^{2}$ is positive and
increases for various values of $m$ and PDE parameters. This shows
the stability of the reconstructed PDE $f(Q)$ model throughout
cosmic evolution.
\begin{figure}
\epsfig{file=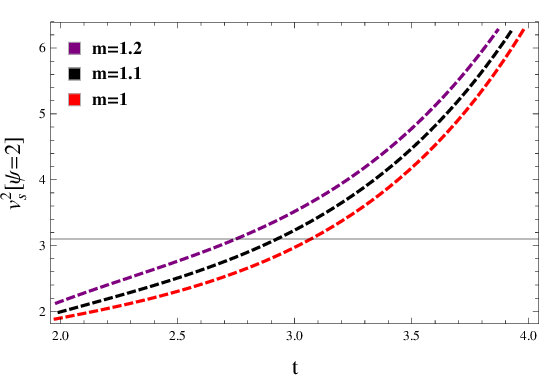,width=.5\linewidth}
\epsfig{file=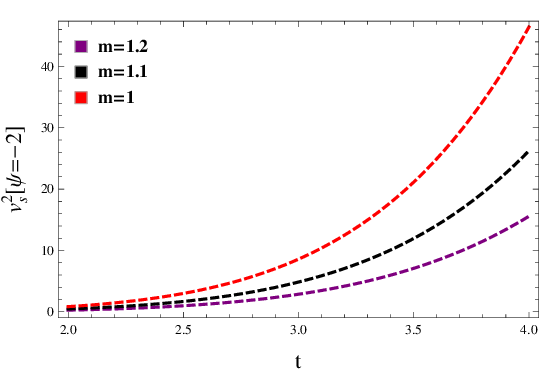,width=.5\linewidth}\caption{Graphs of
$\nu_s^{2}$ ($\psi=\pm2$) with $t$.}
\end{figure}

\section{Reconstruction of GGPDE $f(Q)$ Model}

Here, we reconstruct the GGPDE $f(Q)$ model through the
corresponding principle. Using Eqs.\eqref{2} and \eqref{19}, we
obtain
\begin{equation}\label{24}
\frac{f}{2}-6H^{2}f_{Q}=(\alpha H+\beta H^{2})^{\psi}.
\end{equation}
This implies the following solution
\begin{equation}\label{25}
f(Q)=\frac{1}{6} \bigg(6 c \sqrt{Q}-\alpha  6^{-\frac{\psi }{2}}
Q^{\frac{\psi }{2}-1} \bigg(2 \sqrt{6} \beta  Q^{3/2}+\frac{12
Q}{\psi -1}\bigg)\bigg).
\end{equation}
This model with a power-law is found by substituting Eq.\eqref{9}
into \eqref{25} as
\begin{equation}\label{26}
f(Q)=\frac{\sqrt{6} c (\psi -1) \sqrt{\frac{m^2}{t^2}}-2 \alpha
\bigg(\frac{m^2}{t^2}\bigg)^{\psi /2} \bigg(\beta  (\psi -1)
\sqrt{\frac{m^2}{t^2}}+1\bigg)}{\psi -1}.
\end{equation}
Figure \textbf{8} shows the behavior of the reconstructed GGPDE
$f(Q)$ gravity model which decreases with increasing values of $t$
for $\psi=2$. On the other hand, for $\psi=-2$, $f(Q)$ increases
with increasing values of $t$. The values of $\rho_D$ and $p_D$ are
found using Eq.\eqref{25} in \eqref{19} and \eqref{20} as
\begin{figure}
\epsfig{file=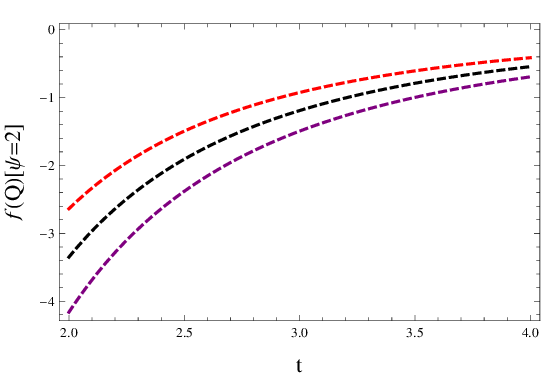,width=0.5\linewidth}
\epsfig{file=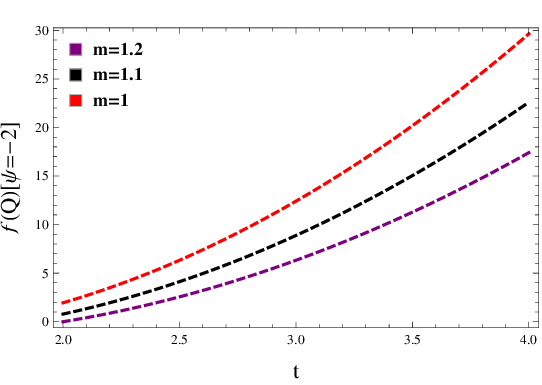,width=0.5\linewidth}\caption{Graphical
representation of $f(Q)$ ($\psi=\pm2$) with $t$.}
\end{figure}
\begin{eqnarray}\nonumber
\rho_D&=&\alpha  6^{-\frac{\psi }{2}-1} Q^{\psi /2} \bigg(\sqrt{6}
\beta  \sqrt{Q} \psi +6\bigg)+c,\\\nonumber p_D&=&\frac{1}{Q^{5/2}
(\psi -1)}\bigg[2^{\frac{1}{2} (-\psi -3)} 3^{-\frac{\psi }{2}-1}
\bigg(c Q^2 2^{\frac{\psi +3}{2}} 3^{\frac{\psi }{2}+1} (\psi -1)
\\\nonumber&\times&\big(\dot{H}+3 H^2\big)-c H Q 2^{\frac{\psi +1}{2}}
3^{\frac{\psi }{2}+1} (\psi -1)+c Q^3 2^{\frac{\psi +1}{2}}
3^{\frac{\psi }{2}+1} \\\nonumber&\times& (\psi -1)-4 \sqrt{3}
\alpha  \beta  \big(\psi ^2-1\big)\big(\dot{H}+3 H^2\big)
Q^{\frac{\psi }{2}+2}-12 \sqrt{2}
\\\nonumber&\times&\alpha \psi \big(\dot{H}+3
H^2\big) Q^{\frac{\psi +3}{2}}-2 \sqrt{3} \alpha \beta H (\psi
-1)^2(\psi +1) Q^{\frac{\psi }{2}+1}\\\nonumber&-&6 \sqrt{2} \alpha
( Q^{\frac{\psi +5}{2}}+H (\psi -2) \psi Q^{\frac{\psi +1}{2}})-2
\sqrt{3} \alpha \beta (\psi -1) Q^{\frac{\psi }{2}+3}\bigg)\bigg].
\end{eqnarray}
Replacing Eq.\eqref{9} into the equations above, we can represent
them in terms of $t$ as
\begin{eqnarray}\nonumber
\rho_D&=&\alpha  \bigg(\frac{m^2}{t^2}\bigg)^{\psi /2} \bigg(\beta
\psi  \sqrt{\frac{m^2}{t^2}}+1\bigg),\\\nonumber p_D&=&\frac{1}{72
\sqrt{3} m^5 (\psi -1)}\bigg[\sqrt{\frac{m^2}{t^2}} \bigg(m^2 t^2
\bigg(24 \sqrt{3} \alpha  \psi
\bigg(\frac{m^2}{t^2}\bigg)^{\frac{\psi +1}{2}}-t
\\\nonumber&\times& (\psi
-1)\bigg(3 \sqrt{2} c+2 \sqrt{3} \alpha  \beta \bigg(\psi ^2-1\bigg)
\bigg(\frac{m^2}{t^2}\bigg)^{\frac{\psi}{2}}\bigg)\bigg)-72 m^5
\\\nonumber&\times& (\psi -1)
\bigg(\sqrt{3} \alpha \beta  (\psi +2)
\bigg(\frac{m^2}{t^2}\bigg)^{\frac{\psi}{2}}-3 \sqrt{2} c\bigg)+12
(\psi -1)\\\nonumber&\times&m^4\bigg(2 \sqrt{3} \alpha \beta  (\psi
+1) \bigg(\frac{m^2}{t^2}\bigg)^{\frac{\psi}{2}}-3 \sqrt{2}
c\bigg)-2 \sqrt{3} \alpha  t^5 (\psi -2) \\\nonumber&\times&\psi
\bigg(\frac{m^2}{t^2}\bigg)^{\frac{\psi +1}{2}}-72 \sqrt{3} \alpha
m^3 t^2 (\psi +1) \bigg(\frac{m^2}{t^2}\bigg)^{\frac{\psi
+1}{2}}\bigg)\bigg].
\end{eqnarray}
Figures \textbf{9} and \textbf{10} illustrate the behavior of $p_D$
and $\rho_D$. In Figure \textbf{9}, the reconstructed GGPDE $f(Q)$
gravity shows an exponentially increasing behavior of $\rho_D$, for
both $\psi=\pm2$. On the other hand, in Figure \textbf{10}, for both
$\psi=\pm2$, $p_D$ follows a negative trend that decreases over
time, consistent with the expected DE behavior.
\begin{figure}
\epsfig{file=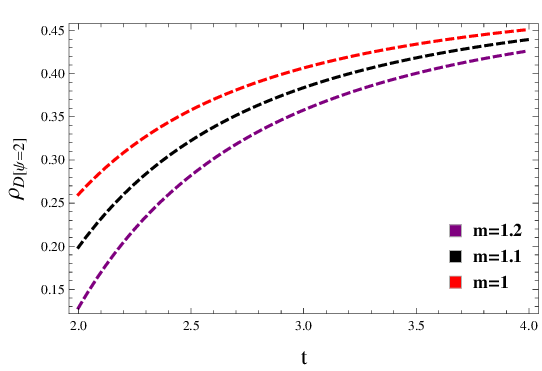,width=.5\linewidth}
\epsfig{file=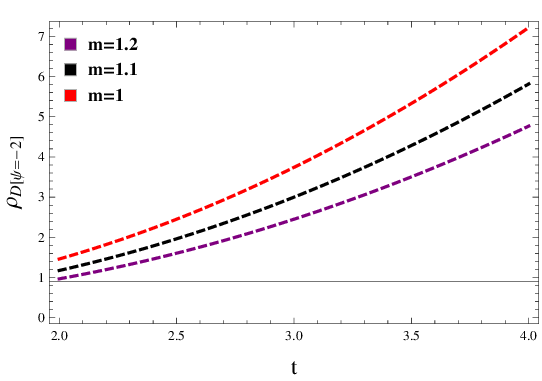,width=.48\linewidth}\caption{Plots of $\rho_D$
($\psi=\pm2$) with $t$.}
\end{figure}
\begin{figure}
\epsfig{file=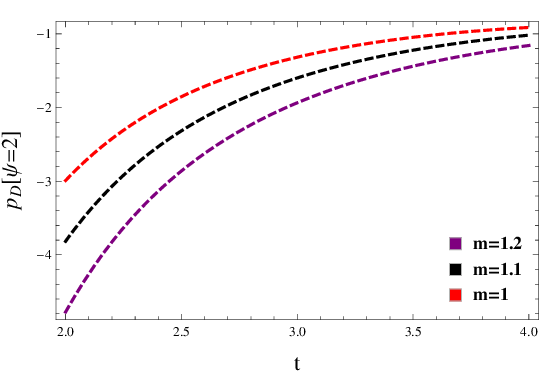,width=.49\linewidth}
\epsfig{file=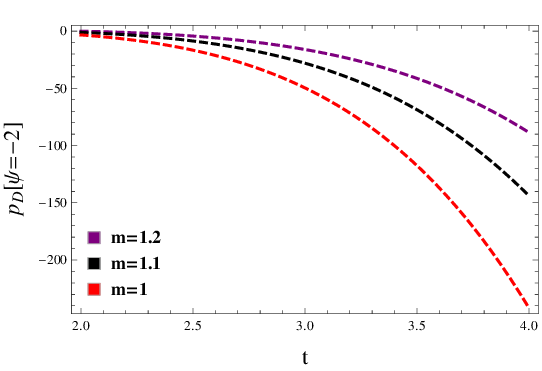,width=.5\linewidth}\caption{Plots of $p_D$
($\psi=\pm2$) against $t$.}
\end{figure}
\begin{figure}
\epsfig{file=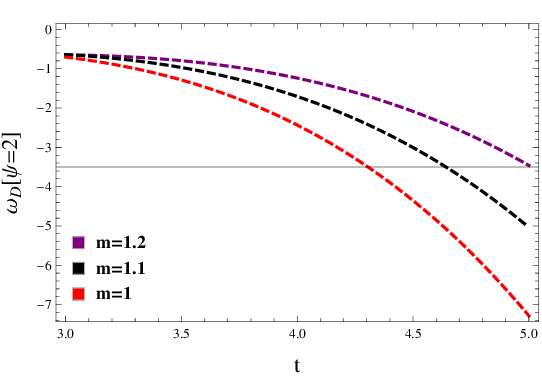,width=.5\linewidth}
\epsfig{file=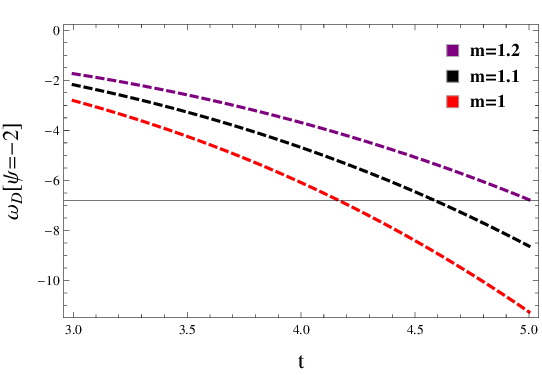,width=.5\linewidth}\caption{Graphs show of
$\omega_{D}$ ($\psi=\pm2$) versus $t$.}
\end{figure}

Now we examine the various phases of the universe evolution. We
illustrate the behavior of different cosmological parameters, such
as the EoS parameter and phase planes as well as stability of this
model. The EoS parameter is a crucial quantity in cosmology that
describes the relationship between the pressure and the energy
density of a given component of the universe. It plays a fundamental
role in characterizing the nature of DE and its impact on the
evolution of the universe. The EoS parameter is given by
\begin{eqnarray}\nonumber
\omega_{D}&=&-\bigg\{\sqrt{\frac{m^2}{t^2}} t^{3 m} \bigg(m^2 t^2
\bigg(t (\psi -1) \bigg(3 \sqrt{2} c+2 \sqrt{3} \alpha  \beta
\bigg(\psi ^2-1\bigg) \bigg(\frac{m^2}{t^2}\bigg)^{\psi
/2}\bigg)\\\nonumber&-&24 \sqrt{3} \alpha  \psi
\bigg(\frac{m^2}{t^2}\bigg)^{\frac{\psi +1}{2}}\bigg)+72 m^5 (\psi
-1) \bigg(\sqrt{3} \alpha  \beta  (\psi +2)
\bigg(\frac{m^2}{t^2}\bigg)^{\frac{\psi}{2}}-3 \sqrt{2}
c\bigg)\\\nonumber&-&12 m^4 (\psi -1) \bigg(2 \sqrt{3} \alpha  \beta
(\psi +1) \bigg(\frac{m^2}{t^2}\bigg)^{\psi /2}-3 \sqrt{2} c\bigg)+2
\sqrt{3} \alpha  t^5 (\psi -2) \psi\\\nonumber&\times&
\bigg(\frac{m^2}{t^2}\bigg)^{\frac{\psi +1}{2}}+72 \sqrt{3} \alpha
m^3 t^2 (\psi +1) \bigg(\frac{m^2}{t^2}\bigg)^{\frac{\psi
+1}{2}}\bigg)\bigg\}\bigg\{72 \sqrt{3} m^5 (\psi -1)
\\\nonumber&\times&\bigg(\alpha t^{3 m}
\bigg(\frac{m^2}{t^2}\bigg)^{\psi /2} \bigg(\beta  \psi
\sqrt{\frac{m^2}{t^2}}+1\bigg)+\xi \bigg)\bigg\}^{-1}.
\end{eqnarray}
Figure \textbf{11} shows that $\omega_{D}<-1$, implying that the
model includes phantom field DE for both $\psi=\pm2$, which aligns
with the PDE phenomenon. The $\omega_{D}-\omega^{\prime}_{D}$-plane
is used in cosmology to analyze the evolution and properties of DE.
This plane can indicate whether a particular DE model evolves
towards the different trajectories such as cosmological constant,
phantom energy scenario or quintessence model. By examining these
trajectories, one can conclude the important characteristics about
the influence of DE on the universe expansion and future evolution.
The value of $\omega^{\prime}_{D}$ is given as
\begin{eqnarray}\nonumber
\omega^{\prime}_{D}&=&\bigg\{t^{3 m} \bigg(72 m^5 t^{3 m} \alpha
\beta \psi  \bigg(\psi ^2-1\bigg) \bigg(2 \sqrt{3}
\bigg(\frac{m^2}{t^2}\bigg)^{\psi /2} \alpha  \beta -3 \sqrt{2}
c\bigg) \bigg(\frac{m^2}{t^2}\bigg)^{\psi /2}\\\nonumber&-&12 m^4
t^{3 m} \alpha \beta  (\psi -1) \psi  \bigg(4 \sqrt{3}
\bigg(\frac{m^2}{t^2}\bigg)^{\psi /2} \alpha  \beta  (\psi +1)-3
\sqrt{2} (\psi +2) c\bigg)
\\\nonumber&\times&\bigg(\frac{m^2}{t^2}\bigg)^{\psi /2}+72 m^3 t^2
\alpha \psi \bigg(t^{3 m} \bigg(2 \sqrt{3}
\bigg(\frac{m^2}{t^2}\bigg)^{\psi /2} \alpha  \bigg(2
\sqrt{\frac{m^2}{t^2}} \beta  \psi +1\bigg)\\\nonumber&-&3 \sqrt{2}
\sqrt{\frac{m^2}{t^2}} (\psi -1) c\bigg)-\sqrt{3} \xi (\psi -1)
\bigg(\sqrt{\frac{m^2}{t^2}} \beta  (\psi +1)+1\bigg)\bigg)
\bigg(\frac{m^2}{t^2}\bigg)^{\psi /2}\\\nonumber&+& \bigg(t^{3 m+1}
\alpha \beta  (\psi -1) \psi  \bigg(8 \sqrt{3} \alpha  \beta
\big(\psi ^2-1\big) \bigg(\frac{m^2}{t^2}\bigg)^{\psi /2}+3 \sqrt{2}
(\psi +4) c\bigg)\\\nonumber&\times&
\bigg(\frac{m^2}{t^2}\bigg)^{\psi /2}+12 \xi \bigg(2 \sqrt{3} \alpha
\bigg(\sqrt{\frac{m^2}{t^2}} \beta (\psi +1) (\psi -1)^2+(\psi -2)
\psi \bigg)
\\\nonumber&\times&\bigg(\frac{m^2}{t^2}\bigg)^{\frac{\psi}{2}}+3
\sqrt{2} (\psi -1) c \sqrt{\frac{m^2}{t^2}}\bigg)+t^{3 m} \bigg(36
\sqrt{2} \bigg(\frac{m^2}{t^2}\bigg)^{\frac{\psi +1}{2}} \alpha
\big(\psi ^2-1\big) c\\\nonumber&-&24 \sqrt{3}
\bigg(\frac{m^2}{t^2}\bigg)^{\psi } \alpha ^2 \bigg(2 \psi
+\sqrt{\frac{m^2}{t^2}} \beta  \big(4 \psi
^2-1\big)\bigg)\bigg)\bigg)m^2 t^2+t^5 \bigg(t^{3 m} \bigg(3
\sqrt{2}\\\nonumber&\times& \alpha  \big(\psi ^2+2 \psi -3\big) c
\bigg(\frac{m^2}{t^2}\bigg)^{\frac{\psi +1}{2}}+2 \sqrt{3} \alpha ^2
\bigg(4 (\psi -2) \psi +\sqrt{\frac{m^2}{t^2}}\big(3+8 \psi
^3\\\nonumber&-&13 \psi ^2-3 \psi \big)\beta\bigg)
\bigg(\frac{m^2}{t^2}\bigg)^{\psi }\bigg)-\xi  \bigg(2 \sqrt{3}
\bigg(\frac{m^2}{t^2}\bigg)^{\frac{\psi}{2}} \alpha
\bigg(\sqrt{\frac{m^2}{t^2}} \beta \big(\psi ^2-2 \psi
-3\big)\\\nonumber&\times& (\psi -1)^2+\psi  \bigg(\psi ^2-6 \psi
+8\bigg)\bigg)-9 \sqrt{2} \sqrt{\frac{m^2}{t^2}} (\psi -1)
c\bigg)\bigg)\bigg)\bigg\}\bigg\{864 \sqrt{3}\\\nonumber&\times& m^5
(\psi -1) \bigg(t^{3 m} \alpha \bigg(\sqrt{\frac{m^2}{t^2}} \beta
\psi +1\bigg) \bigg(\frac{m^2}{t^2}\bigg)^{\psi /2}+\xi
\bigg)^2\bigg\}^{-1}.
\end{eqnarray}
Figure \textbf{12} demonstrates that
$\omega_{D}<0,~\omega^{\prime}_{D}<0$ for all values of $m$ and PDE
parameter, indicating the existence of a freezing region.
\begin{figure}
\epsfig{file=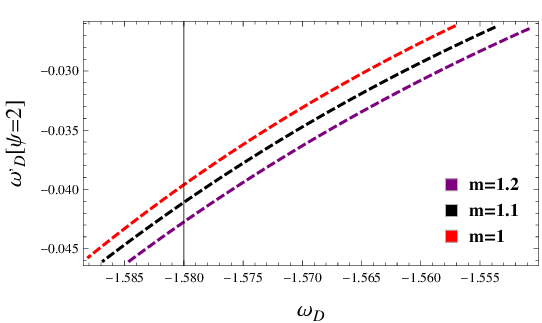,width=.55\linewidth}
\epsfig{file=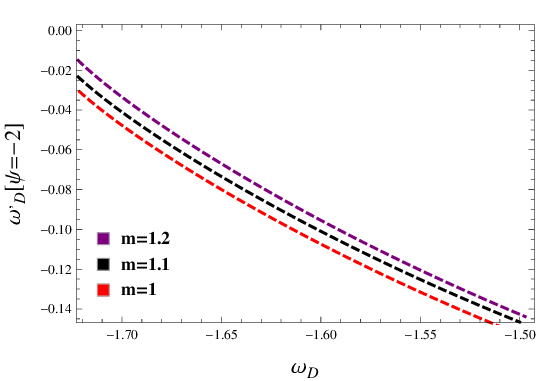,width=.47\linewidth}\caption{Plot of
$\omega^{\prime}_{D}$ ($\psi=\pm2$) versus $\omega_{D}$.}
\end{figure}
The $(r-s)$-plane is a powerful tool in cosmology for
differentiating between various DE models and understanding the
cosmological dynamics. It uses a combination of the Hubble parameter
and its time derivatives to analyze cosmic expansion. Specifically,
the parameter $s$ describes the acceleration of cosmic expansion,
while the parameter $r$ shows deviations from pure power-law
behavior. It can be used to distinguish between different DE
scenarios like Chaplygin gas, Holographic DE, standard cold DM and
quintessence without relying on any particular model. The values of
$r$ and $s$ are given in Appendix \textbf{B}. Figure \textbf{13}
shows the Chaplygin gas model ($r>1,~s<0$) for various values of $m$
and PDE parameter.
\begin{figure}
\epsfig{file=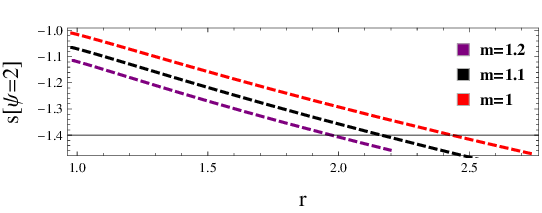,width=.49\linewidth}
\epsfig{file=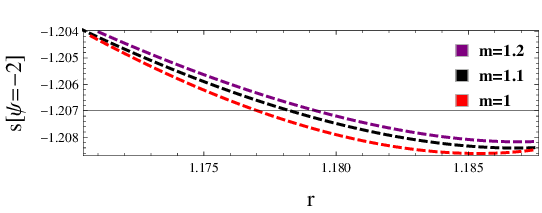,width=.5\linewidth}\caption{Plots illustrate
$s$ ($\psi=\pm2$) with $r$.}
\end{figure}
In cosmology, especially in DE models, $\nu_{s}^{2}$ helps to
determine the stability of the universe expansion. A positive value
indicates a stable configuration, while a negative value suggests
unstable behavior for the corresponding model. Thus, the squared
speed of sound is essential for understanding the dynamics of
various components in the universe. The value of $\nu_{s}^{2}$ turns
out to be
\begin{eqnarray}\nonumber
\nu_{s}^{2}&=&-\bigg\{\bigg(\frac{m^2}{t^2}\bigg)^{\frac{1}{2}-\frac{\psi
}{2}} t^{3 m} \bigg(2 \sqrt{3} t^5 \alpha  (\psi -2) \psi
\bigg(\frac{m^2}{t^2}\bigg)^{\frac{\psi +1}{2}}+72 \sqrt{3} m^3 t^2
\alpha  (\psi +1)\\\nonumber&\times&
\bigg(\frac{m^2}{t^2}\bigg)^{\frac{\psi +1}{2}}-12 m^4 (\psi -1)
\bigg(2 \sqrt{3} \bigg(\frac{m^2}{t^2}\bigg)^{\psi /2} \alpha  \beta
(\psi +1)-3 \sqrt{2} c\bigg)+72 \\\nonumber&\times& m^5 (\psi -1)
\bigg(\sqrt{3} \bigg(\frac{m^2}{t^2}\bigg)^{\psi /2} \alpha  \beta
(\psi +2)-3 \sqrt{2} c\bigg)+ \bigg(t (\psi -1) \bigg(2 \sqrt{3}
\alpha \beta \\\nonumber&\times& \big(\psi ^2-1\big)
\bigg(\frac{m^2}{t^2}\bigg)^{\psi /2}+3 \sqrt{2} c\bigg)-24 \sqrt{3}
\bigg(\frac{m^2}{t^2}\bigg)^{\frac{\psi +1}{2}} \alpha \psi
\bigg)m^2 t^2\bigg) \bigg(36 m^5 t^{3 m} \\\nonumber&\times&\alpha
\psi \bigg(\sqrt{6} \bigg(\frac{m^2}{t^2}\bigg)^{\psi /2} \alpha
\beta \bigg(2 \psi ^2+3 \psi +\sqrt{\frac{m^2}{t^2}} \beta
\bigg(\psi ^3+2 \psi ^2-\psi -2\bigg)-1\bigg)\\\nonumber&-&6 (\psi
-1) \bigg(\sqrt{\frac{m^2}{t^2}} \beta  (\psi +1)+1\bigg) c\bigg)
\bigg(\frac{m^2}{t^2}\bigg)^{\psi /2}+\sqrt{6} \bigg(4
\bigg(\frac{m^2}{t^2}\bigg)^{\psi /2} t^{3 m} \alpha
\\\nonumber&-&\xi (\psi -4)\bigg)t^5 \alpha (\psi -2) \psi
\bigg(\frac{m^2}{t^2}\bigg)^{\frac{\psi +1}{2}}+36 \sqrt{6} m^5 t^{3
m} \alpha ^2 \psi  (\psi +1) \bigg(\frac{m^2}{t^2}\bigg)^{\psi
-\frac{1}{2}}\\\nonumber&+&m^2 t^2 \bigg(t^{3 m+1} \alpha
\bigg(\sqrt{6} \alpha \beta  \bigg(4 \sqrt{\frac{m^2}{t^2}} \beta
\psi ^4+\bigg(8-4 \sqrt{\frac{m^2}{t^2}} \beta \bigg) \psi
^3-\bigg(4 \sqrt{\frac{m^2}{t^2}} \beta \\\nonumber&+&13\bigg) \psi
^2+\bigg(4 \sqrt{\frac{m^2}{t^2}} \beta -3\bigg) \psi +3\bigg)
\bigg(\frac{m^2}{t^2}\bigg)^{\psi /2}+3 (\psi -1)
\bigg(\sqrt{\frac{m^2}{t^2}} \beta  \psi ^2\\\nonumber&+&4
\sqrt{\frac{m^2}{t^2}} \beta  \psi +\psi +3\bigg) c\bigg)
\bigg(\frac{m^2}{t^2}\bigg)^{\psi /2}+12 \sqrt{6} \alpha  \xi (\psi
-2) \psi \bigg(\frac{m^2}{t^2}\bigg)^{\frac{\psi
+1}{2}}\\\nonumber&-&24 \sqrt{6} t^{3 m} \alpha ^2 \psi
\bigg(\frac{m^2}{t^2}\bigg)^{\psi +\frac{1}{2}}-t \xi  (\psi -1)
\bigg( \alpha  \beta \big(\psi ^3-3 \psi ^2-\psi +3\big)\sqrt{6}
\\\nonumber&\times&\bigg(\frac{m^2}{t^2}\bigg)^{\psi /2}-9 c\bigg)\bigg)-12 m^4
\bigg(\bigg(\frac{m^2}{t^2}\bigg)^{\frac{\psi}{2}} t^{3 m} \alpha
\bigg(\sqrt{6} \bigg(\frac{m^2}{t^2}\bigg)^{\frac{\psi}{2}} \alpha
\beta \bigg(4 \psi ^2+2\beta\\\nonumber&\times&
\sqrt{\frac{m^2}{t^2}} \big(\psi ^2-1\big) \psi -1\bigg)-3 (\psi -1)
\bigg(\sqrt{\frac{m^2}{t^2}} \beta  \psi ^2+2 \sqrt{\frac{m^2}{t^2}}
\beta  \psi +\psi +1\bigg) c\bigg)\\\nonumber&-&\xi (\psi -1)
\bigg(\sqrt{6} \alpha  \beta  \big(\psi ^2-1\big)
\bigg(\frac{m^2}{t^2}\bigg)^{\frac{\psi}{2}}+3
c\bigg)\bigg)\bigg)\bigg\}\bigg\{7776 \sqrt{2} m^8 \alpha (\psi
-1)^2\\\nonumber&\times&  \psi \bigg(\beta  (\psi +1) m^2+t^2
\sqrt{\frac{m^2}{t^2}}\bigg) \bigg(t^{3 m} \alpha
\bigg(\sqrt{\frac{m^2}{t^2}} \beta  \psi +1\bigg)
\bigg(\frac{m^2}{t^2}\bigg)^{\frac{\psi}{2}}+\xi
\bigg)^2\bigg\}^{-1}.
\end{eqnarray}
Figure \textbf{14} indicates that  $\nu_{s}^{2}$ is positively
increasing. This observation implies that the ongoing phase of
cosmic expansion aligns well with the stable non-interacting GGPDE
model.
\begin{figure}
\epsfig{file=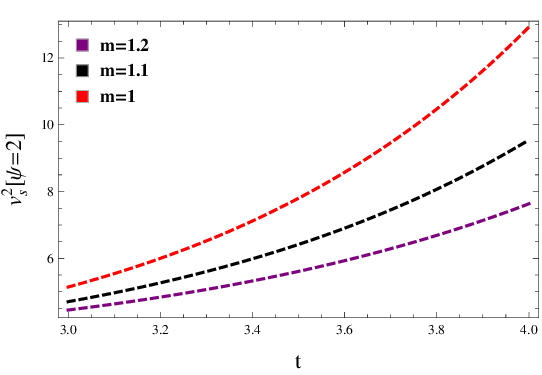,width=.5\linewidth}
\epsfig{file=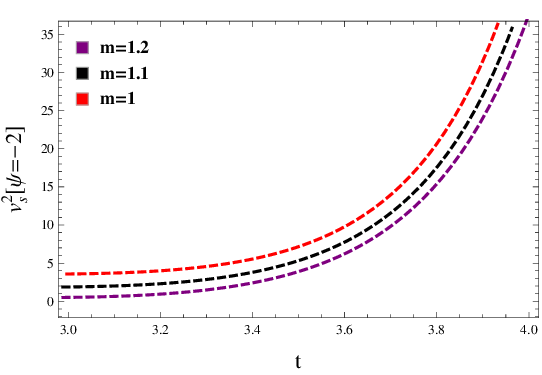,width=.5\linewidth}\caption{Graphs of
$\nu_s^{2}$ ($\psi=\pm2$) versus $t$.}
\end{figure}

\section{Conclusions}

In this paper, we have explored two different DE models in the
context of $f(Q)$ gravity. We have reconstructed PDE and GGPDE
$f(Q)$ gravity models through a correspondence scheme for the FRW
model with a power-law scale factor to examine the non-interacting
case. We have then analyzed the evolutionary trajectories of the EoS
as well as the $\omega_{D}-\omega^{\prime}_{D}$ and the
$(r-s)$-planes. Finally, we have explored the stability of the DE
model. The summary of the results obtained are presented as follows.
\begin{itemize}
\item
Both DE models exhibit a decreasing pattern when $\psi =2$, but an
increasing trend when $\psi=-2$, suggesting that the reconstructed
model is realistic (Figures \textbf{1} and \textbf{8}).
\item
In both models, the energy density increases and pressure decreases
for all values of $\psi$ and $m$. This behavior aligns with the
characteristics of DE (Figures \textbf{2}, \textbf{9} and
\textbf{3}, \textbf{10}). For PDE and GGPDE models, the increasing
energy density and decreasing pressure indicate that DE becomes more
dominant as the universe evolves. This is crucial for the
accelerated expansion observed in the universe today. Their
alignment with the fundamental properties of DE supports their
viability as explanations for one of cosmology's most significant
observations. We have plotted the energy density and pressure and
their behavior is consistent with \cite{1a}.
\item
We have found that the EoS parameter shows phantom region as the
values of $m$ decreases (Figures \textbf{4} and \textbf{11}).
Consequently, our findings support the current accelerated expansion
of the universe. Planck 2015 measurements have suggested alternative
values for $\omega_{D}$ with a confidence level of 95\% \cite{54}
\begin{eqnarray}\nonumber
\omega_{D}&=&-1.023^{+0.091}_{-0.096}\quad(\text{Planck
TT+LowP+ext}),\\\nonumber
\omega_{D}&=&-1.006^{+0.085}_{-0.091}\quad(\text{Planck
TT+LowP+lensing+ext}),\\\nonumber
\omega_{D}&=&-1.019^{+0.075}_{-0.080}\quad (\text{Planck TT, TE,
EE+LowP+ext}).
\end{eqnarray}
It is important to highlight that in non-interacting universe
models, the effective EoS parameter is found to be consistent with
observable constraints. Notably, when comparing to the GGPDE model,
the EoS parameter in the PDE model proves to be more compatible with
observational data.
\item
The $\omega_{D}-\omega^{\prime}_{D}$-plane illustrates the freezing
region for all values of $m$ and the PDE parameter (Figures
\textbf{5} and \textbf{12}). This shows that the universe appears to
undergo a more rapid expansion.
\item
The $(r-s)$-plane behaves like the Chaplygin gas model in the
non-interacting case for different values of $m$ and PDE parameters
(Figures \textbf{6} and \textbf{13}).
\item
The behavior of $\nu_{s}^{2}$ remains positive for all values of
$m$, indicating stability (Figures \textbf{7} and \textbf{14}).
\end{itemize}
The cosmography of PDE and GGPDE within the $f(Q)$ framework
surpasses other modified gravity theories. The cosmography of PDE
and GGPDE in $f(Q)$ gravity provides a more accurate, stable and
consistent framework for understanding DE \cite{21}-\cite{21b}. Thus
we can conclude that our results are more accurate in descriptions
of the cosmic accelerated expansion.

\section*{Appendix A: Calculation of $r$ and $s$ in PDE}
\renewcommand{\theequation}{A\arabic{equation}}
\setcounter{equation}{0}
\begin{eqnarray}\nonumber
r&=&\frac{1}{m^{12} \bigg(t^{3 m} \alpha
\bigg(\frac{m^2}{t^2}\bigg)^{\frac{\psi}{2}}+\xi \bigg)^4 (\psi
-1)^2}\bigg[2^{-2 \psi -11} 3^{-2 (\psi +3)} t^{12} \bigg(-6^{2 \psi
+3}  t^{3 m-7}\\\nonumber&\times& m^4\bigg(t^{3 m} \alpha
\bigg(\frac{m^2}{t^2}\bigg)^{\psi /2}+\xi \bigg) (\psi -1) \bigg(-72
\bigg(-t^{3 m} \alpha  \xi \bigg(\big( \psi ^2+7 \psi
-14\big)\psi\\\nonumber&\times&\bigg(\frac{m^2}{t^2}\bigg)^{\psi /2}
\alpha-\sqrt{6} \sqrt{\frac{m^2}{t^2}} \bigg(\psi ^3-3 \psi
^2+2\bigg) c\bigg) \bigg(\frac{m^2}{t^2}\bigg)^{\psi /2}+t^{6 m}
\alpha ^2 \bigg(8\alpha \psi\\\nonumber&\times&
\bigg(\frac{m^2}{t^2}\bigg)^{\psi /2} -\sqrt{6}
\sqrt{\frac{m^2}{t^2}} (\psi -1) (\psi +1)^2 c\bigg)
\bigg(\frac{m^2}{t^2}\bigg)^{\psi }+\xi ^2
\bigg(\bigg(\frac{m^2}{t^2}\bigg)^{\psi /2} \alpha
\psi\\\nonumber&\times& \big(\psi ^2-5 \psi +6\big)-\sqrt{6}
\sqrt{\frac{m^2}{t^2}} (\psi -1) c\bigg)\bigg) m^3+12 \bigg(-t^{3 m}
\alpha  \xi  \bigg(2 \bigg(\frac{m^2}{t^2}\bigg)^{\psi /2}
\\\nonumber&\times&\alpha \psi  \big(\psi ^2+6 \psi
-16\big)-\sqrt{6} \sqrt{\frac{m^2}{t^2}} \big(\psi ^3-5 \psi ^2-2
\psi +6\big) c\bigg) \bigg(\frac{m^2}{t^2}\bigg)^{\psi /2}+t^{6
m}\\\nonumber&\times& \alpha ^2 \bigg(16 \alpha  \psi
\bigg(\frac{m^2}{t^2}\bigg)^{\psi /2}+\sqrt{6} \big(3-\psi ^3-3 \psi
^2+\psi\big) c \sqrt{\frac{m^2}{t^2}}\bigg)
\bigg(\frac{m^2}{t^2}\bigg)^{\psi }+\xi ^2\\\nonumber&\times&
\bigg(2 \bigg(\frac{m^2}{t^2}\bigg)^{\psi /2} \alpha  \psi \big(\psi
^2-6 \psi +8\big)-3 \sqrt{6} \sqrt{\frac{m^2}{t^2}} (\psi -1)
c\bigg)\bigg) m^2+t^3 \bigg(t^{3 m}\\\nonumber&\times& \alpha  \xi
\bigg(2 \alpha \psi \big(\psi ^3+8 \psi ^2-68 \psi +96\big)
\bigg(\frac{m^2}{t^2}\bigg)^{\psi /2}+ \big(\psi ^3-9 \psi ^2-22
\psi +30\big) \\\nonumber&\times&\sqrt{6}c
\sqrt{\frac{m^2}{t^2}}\bigg) \bigg(\frac{m^2}{t^2}\bigg)^{\psi
/2}-t^{6 m} \alpha ^2 \bigg(48 \alpha  (\psi -2) \psi
\bigg(\frac{m^2}{t^2}\bigg)^{\psi /2}+\sqrt{6} \bigg(\psi ^3+7
\psi\\\nonumber&+&7 \psi ^2 -15\bigg) c \sqrt{\frac{m^2}{t^2}}\bigg)
\bigg(\frac{m^2}{t^2}\bigg)^{\psi }-\xi ^2 \bigg(2 \alpha  \psi
\big(\psi ^3-12 \psi ^2+44 \psi -48\big)
\bigg(\frac{m^2}{t^2}\bigg)^{\psi /2}\\\nonumber&+&15 \sqrt{6} (\psi
-1) c \sqrt{\frac{m^2}{t^2}}\bigg)\bigg)\bigg)+\bigg[2^{2 \psi +5}
9^{\psi +2} m \bigg(\frac{m^2}{t^2}\bigg)^{\frac{5}{2}} \bigg(t^{3
m} \alpha \bigg(\frac{m^2}{t^2}\bigg)^{\frac{\psi}{2}}+\xi \bigg)^2
\\\nonumber&\times&(\psi -1) \bigg(2 t^{3 m+5} \alpha  \bigg(4
\bigg(\frac{m^2}{t^2}\bigg)^{\psi /2} t^{3 m} \alpha -\xi  (\psi
-4)\bigg) (\psi -2) \psi \bigg(\frac{m^2}{t^2}\bigg)^{\frac{\psi
+1}{2}}\\\nonumber&+&72 m^3 t^{3 m+2} \alpha  \psi  \bigg(2 t^{3 m}
\alpha \bigg(\frac{m^2}{t^2}\bigg)^{\psi /2}+\xi -\xi  \psi \bigg)
\bigg(\frac{m^2}{t^2}\bigg)^{\frac{\psi +1}{2}}+12 \sqrt{6} m^4 t^{3
m}\\\nonumber&\times& (\psi -1) \bigg(t^{3 m} \alpha  (\psi +1)
\bigg(\frac{m^2}{t^2}\bigg)^{\psi /2}+\xi \bigg) c+m^2 t^{3 m+2}
\bigg(\sqrt{6} t^{3 m+1} \alpha \bigg(\frac{m^2}{t^2}\bigg)^{\psi
/2} \\\nonumber&\times&\bigg(\psi ^2+2 \psi -3\bigg) c +24 \alpha
\xi  (\psi -2) \psi \bigg(\frac{m^2}{t^2}\bigg)^{\frac{\psi
+1}{2}}-48 t^{3 m} \alpha ^2 \psi  \bigg(\frac{m^2}{t^2}\bigg)^{\psi
+\frac{1}{2}}\\\nonumber&+&3 \sqrt{6} t \xi  (\psi -1) c\bigg)+72
m^5 (\psi -1) \bigg(8 t^{3 m} \alpha \xi
\bigg(\frac{m^2}{t^2}\bigg)^{\frac{\psi +1}{2}}+4 \xi ^2
\sqrt{\frac{m^2}{t^2}}+t^{6 m}\\\nonumber&\times& \bigg(4
\bigg(\frac{m^2}{t^2}\bigg)^{\psi +\frac{1}{2}} \alpha ^2-\sqrt{6}
\bigg(\frac{m^2}{t^2}\bigg)^{\psi /2} \alpha  \psi
c\bigg)\bigg)\bigg)\bigg]\frac{1}{t^6}+\frac{1}{t^{10}}\bigg[4^{\psi
} 9^{\psi +1} \bigg(2 t^{3 m+5} \alpha \psi\\\nonumber&\times&
\bigg(4 \bigg(\frac{m^2}{t^2}\bigg)^{\psi /2} t^{3 m} \alpha -\xi
(\psi -4)\bigg) (\psi -2)  \bigg(\frac{m^2}{t^2}\bigg)^{\frac{\psi
+1}{2}}+72 m^3 t^{3 m+2} \alpha  \psi\bigg(2 \alpha
\\\nonumber&\times&t^{3 m}\bigg(\frac{m^2}{t^2}\bigg)^{\psi
/2}+\xi -\xi  \psi \bigg) \bigg(\frac{m^2}{t^2}\bigg)^{\frac{\psi
+1}{2}}+12 \sqrt{6} m^4 t^{3 m} (\psi -1) \bigg(t^{3 m}\alpha  (\psi
+1)
\\\nonumber&\times&
\bigg(\frac{m^2}{t^2}\bigg)^{\psi /2}+\xi \bigg) c+m^2 t^{3 m+2}
\bigg(\sqrt{6} t^{3 m+1} \alpha  \big(\psi ^2+2 \psi -3\big) c
\bigg(\frac{m^2}{t^2}\bigg)^{\psi /2}+24\\\nonumber&\times& \alpha
\xi  (\psi -2) \psi \bigg(\frac{m^2}{t^2}\bigg)^{\frac{\psi
+1}{2}}-48 t^{3 m} \alpha ^2 \psi  \bigg(\frac{m^2}{t^2}\bigg)^{\psi
+\frac{1}{2}}+3 \sqrt{6} t \xi  (\psi -1)
c\bigg)+72\\\nonumber&\times& m^5 (\psi -1) \bigg(8 t^{3 m} \alpha
\xi \bigg(\frac{m^2}{t^2}\bigg)^{\frac{\psi +1}{2}}+4 \xi ^2
\sqrt{\frac{m^2}{t^2}}+t^{6 m} \bigg(4
\bigg(\frac{m^2}{t^2}\bigg)^{\psi +\frac{1}{2}} \alpha
^2\\\nonumber&-&\sqrt{6} \bigg(\frac{m^2}{t^2}\bigg)^{\psi /2}
\alpha  \psi c\bigg)\bigg)\bigg)^2\bigg]\bigg)\bigg],
\\\nonumber
s&=&\bigg\{192 m^5 t^{-3 m-2} \bigg(t^{3 m} \alpha
\bigg(\frac{m^2}{t^2}\bigg)^{\psi /2}+\xi \bigg)^2 (\psi -1)
\bigg(\bigg\{2^{-2 \psi -11} 3^{-2 (\psi +3)}
t^{12}\\\nonumber&\times& \bigg(-6^{2 \psi +3} m^4 t^{3 m-7}
\bigg(t^{3 m} \alpha \bigg(\frac{m^2}{t^2}\bigg)^{\psi /2}+\xi
\bigg) (\psi -1) \bigg(-72m^3 \bigg(-t^{3 m} \alpha  \xi
\\\nonumber&\times&\bigg(\alpha\bigg(\frac{m^2}{t^2}\bigg)^{\psi /2}
\psi \big(\psi ^2+7 \psi -14\big)-\sqrt{6} \sqrt{\frac{m^2}{t^2}}
\big(\psi ^3-3 \psi ^2+2\big) c\bigg)
\bigg(\frac{m^2}{t^2}\bigg)^{\psi /2}\\\nonumber&+&t^{6 m} \alpha ^2
\bigg(8 \bigg(\frac{m^2}{t^2}\bigg)^{\psi /2} \alpha \psi -\sqrt{6}
\sqrt{\frac{m^2}{t^2}} (\psi -1) (\psi +1)^2 c\bigg)
\bigg(\frac{m^2}{t^2}\bigg)^{\psi}+\xi ^2
\\\nonumber&\times&\bigg(\bigg(\frac{m^2}{t^2}\bigg)^{\psi /2} \alpha  \psi
\big(\psi^2-5 \psi +6\big)-\sqrt{6} \sqrt{\frac{m^2}{t^2}} (\psi -1)
c\bigg)\bigg) +12 \bigg(-t^{3 m} \alpha\\\nonumber&\times&  \xi
\bigg(2 \big(\frac{m^2}{t^2}\big)^{\psi /2} \alpha  \psi \bigg(\psi
^2+6 \psi -16\bigg)-\sqrt{6} \sqrt{\frac{m^2}{t^2}} \bigg(\psi ^3-5
\psi ^2-2 \psi +6\bigg)
c\bigg)\\\nonumber&\times&\bigg(\frac{m^2}{t^2}\bigg)^{\frac{\psi}{2}}+t^{6
m} \alpha ^2 \bigg(16 \alpha  \psi
\bigg(\frac{m^2}{t^2}\bigg)^{\frac{\psi}{2}}+\sqrt{6} \big(-\psi
^3-3 \psi ^2+\psi +3\big) c \sqrt{\frac{m^2}{t^2}}\bigg)
\\\nonumber&\times&\bigg(\frac{m^2}{t^2}\bigg)^{\psi }+\xi ^2
\bigg(2 \bigg(\frac{m^2}{t^2}\bigg)^{\psi /2} \alpha  \psi
\bigg(\psi ^2-6 \psi +8\bigg)-3 \sqrt{6} \sqrt{\frac{m^2}{t^2}}
(\psi -1) c\bigg)\bigg) \\\nonumber&\times&m^2+t^3 \bigg(t^{3 m}
\alpha  \xi  \bigg(2 \alpha \psi \big(\psi ^3+8 \psi ^2-68 \psi
+96\big) \bigg(\frac{m^2}{t^2}\bigg)^{\psi /2}+\sqrt{6} \big(\psi
^3-9 \psi ^2\\\nonumber&-&22 \psi +30\big) c
\sqrt{\frac{m^2}{t^2}}\bigg) \bigg(\frac{m^2}{t^2}\bigg)^{\psi
/2}-t^{6 m} \alpha ^2 \bigg(48 \alpha  (\psi -2) \psi
\bigg(\frac{m^2}{t^2}\bigg)^{\psi /2}+\sqrt{6} c\\\nonumber&\times&
\big(\psi ^3+7 \psi ^2+7 \psi -15\big) \sqrt{\frac{m^2}{t^2}}\bigg)
\bigg(\frac{m^2}{t^2}\bigg)^{\psi }-\bigg(2 \alpha  \psi \big(\psi
^3-12 \psi ^2+44 \psi -48\big)\\\nonumber&\times&
\bigg(\frac{m^2}{t^2}\bigg)^{\psi /2}+15 \sqrt{6} (\psi -1) c
\sqrt{\frac{m^2}{t^2}}\bigg)\xi ^2 \bigg)\bigg)+\bigg[2^{2 \psi +5}m
9^{\psi +2}(\psi -1) \bigg(\frac{m^2}{t^2}\bigg)^{5/2}
\\\nonumber&\times&\bigg( \alpha
\bigg(\frac{m^2}{t^2}\bigg)^{\psi /2}t^{3 m}+\xi \bigg)^2  \bigg(2
t^{3 m+5} \alpha \bigg(4 \bigg(\frac{m^2}{t^2}\bigg)^{\psi /2} t^{3
m} \alpha -\xi (\psi -4)\bigg) (\psi -2) \\\nonumber&\times&\psi
\bigg(\frac{m^2}{t^2}\bigg)^{\frac{\psi +1}{2}}+72 m^3 t^{3 m+2}
\alpha  \psi  \bigg(2 t^{3 m} \alpha
\bigg(\frac{m^2}{t^2}\bigg)^{\psi /2}+\xi -\xi  \psi \bigg)
\bigg(\frac{m^2}{t^2}\bigg)^{\frac{\psi +1}{2}}
\\\nonumber&+&12m^4 \sqrt{6}t^{3 m} (\psi -1) \bigg(t^{3 m} \alpha  (\psi
+1) \bigg(\frac{m^2}{t^2}\bigg)^{\psi /2}+\xi \bigg) c+m^2 t^{3 m+2}
\bigg(\sqrt{6}c\alpha \\\nonumber&\times&t^{3 m+1}\big(\psi ^2+2
\psi -3\big)  \bigg(\frac{m^2}{t^2}\bigg)^{\frac{\psi}{2}}+24 \alpha
\xi (\psi -2) \psi \bigg(\frac{m^2}{t^2}\bigg)^{\frac{\psi
+1}{2}}-48 t^{3 m} \alpha ^2 \psi
\\\nonumber&\times&\bigg(\frac{m^2}{t^2}\bigg)^{\psi +\frac{1}{2}}+3
\sqrt{6} t \xi  (\psi -1) c\bigg)+72 m^5 (\psi -1) \bigg(8 t^{3 m}
\alpha \xi \bigg(\frac{m^2}{t^2}\bigg)^{\frac{\psi
+1}{2}}\\\nonumber&+&4 \xi ^2 \sqrt{\frac{m^2}{t^2}}+t^{6 m} \bigg(4
\bigg(\frac{m^2}{t^2}\bigg)^{\psi +\frac{1}{2}} \alpha ^2-\sqrt{6}
\bigg(\frac{m^2}{t^2}\bigg)^{\psi /2} \alpha  \psi
c\bigg)\bigg)\bigg)\bigg]\frac{1}{t^6}+\bigg[4^{\psi } 9^{\psi +1}
\\\nonumber&\times&\bigg(2 t^{3 m+5} \alpha  \bigg(4 \bigg(\frac{m^2}{t^2}\bigg)^{\psi
/2} t^{3 m} \alpha -\xi  (\psi -4)\bigg) (\psi -2) \psi
\bigg(\frac{m^2}{t^2}\bigg)^{\frac{\psi +1}{2}}+72
m^3\\\nonumber&\times& t^{3 m+2} \alpha  \psi  \bigg(2 t^{3 m}
\alpha \bigg(\frac{m^2}{t^2}\bigg)^{\psi /2}+\xi -\xi  \psi \bigg)
\bigg(\frac{m^2}{t^2}\bigg)^{\frac{\psi +1}{2}}+12 \sqrt{6} m^4 t^{3
m} (\psi -1) \\\nonumber&\times&\bigg(t^{3 m} \alpha  (\psi +1)
\bigg(\frac{m^2}{t^2}\bigg)^{\psi /2}+\xi \bigg) c+m^2 t^{3 m+2}
\bigg(\sqrt{6} t^{3 m+1} \alpha  \bigg(\psi ^2+2 \psi
-3\bigg)\\\nonumber&\times& c \bigg(\frac{m^2}{t^2}\bigg)^{\psi
/2}+24 \alpha  \xi  (\psi -2) \psi
\bigg(\frac{m^2}{t^2}\bigg)^{\frac{\psi +1}{2}}-48 t^{3 m} \alpha ^2
\psi  \bigg(\frac{m^2}{t^2}\bigg)^{\psi +\frac{1}{2}}+3 \sqrt{6} t
\\\nonumber&\times&\xi  (\psi -1) c\bigg)+72 m^5 (\psi -1)
\bigg(8 t^{3 m} \alpha \xi \bigg(\frac{m^2}{t^2}\bigg)^{\frac{\psi
+1}{2}}+4 \xi ^2 \sqrt{\frac{m^2}{t^2}}+t^{6 m} \bigg(4\alpha
^2\\\nonumber&\times& \bigg(\frac{m^2}{t^2}\bigg)^{\psi
+\frac{1}{2}} -\sqrt{6} \bigg(\frac{m^2}{t^2}\bigg)^{\frac{\psi}{2}}
\alpha \psi
c\bigg)\bigg)\bigg)^2\bigg]\frac{1}{t^{10}}\bigg)\bigg\}\bigg\{m^{12}
\bigg(t^{3 m} \alpha
\bigg(\frac{m^2}{t^2}\bigg)^{\frac{\psi}{2}}+\xi
\bigg)^4\\\nonumber&\times&(\psi
-1)^2\bigg\}^{-1}-1\bigg)\bigg\}\bigg\{72 m^3 \alpha \psi
\bigg(\bigg(2 \bigg(\frac{m^2}{t^2}\bigg)^{\frac{\psi}{2}} \alpha
-\sqrt{6} \sqrt{\frac{m^2}{t^2}} (\psi -1) c\bigg) t^{3
m}\\\nonumber&+&\xi -\xi \psi \bigg)
\bigg(\frac{m^2}{t^2}\bigg)^{\psi /2}+t^3
\bigg(\bigg(\frac{m^2}{t^2}\bigg)^{\psi /2} t^{3 m} \alpha  \bigg(8
\alpha  (\psi -2) \psi  \bigg(\frac{m^2}{t^2}\bigg)^{\psi
/2}+\sqrt{6} \\\nonumber&\times&\bigg(\psi ^2+2 \psi -3\bigg) c
\sqrt{\frac{m^2}{t^2}}\bigg)-\xi  \bigg(2
\bigg(\frac{m^2}{t^2}\bigg)^{\psi /2} \alpha  \psi  \bigg(\psi ^2-6
\psi +8\bigg)-3 \sqrt{6} \\\nonumber&\times&\sqrt{\frac{m^2}{t^2}}
(\psi -1) c\bigg)\bigg)-12 m^2 \bigg(\bigg(4
\bigg(\frac{m^2}{t^2}\bigg)^{\psi } \alpha ^2 \psi -\sqrt{6}
\bigg(\frac{m^2}{t^2}\bigg)^{\frac{\psi +1}{2}} \alpha  \big(\psi
^2-1\big) c\bigg) \\\nonumber&\times&t^{3 m}+\xi \bigg(-2 \alpha
(\psi -2) \psi \bigg(\frac{m^2}{t^2}\bigg)^{\psi /2}-\sqrt{6} (\psi
-1) c \sqrt{\frac{m^2}{t^2}}\bigg)\bigg)\bigg\}^{-1}.
\end{eqnarray}

\section*{Appendix B: Calculation of $r$ and $s$ in GGPDE}
\renewcommand{\theequation}{B\arabic{equation}}
\setcounter{equation}{0}
\begin{eqnarray}\nonumber
r&=&\bigg\{\bigg(288   \beta  (\psi -1) \psi \bigg(t^{3 m} \alpha
\bigg(\sqrt{\frac{m^2}{t^2}} \beta  \psi +2\bigg)
\bigg(\frac{m^2}{t^2}\bigg)^{\frac{\psi}{2}}+2 \xi \bigg)t^{3 m}
\bigg(\frac{m^2}{t^2}\bigg)^{\frac{\psi}{2}}\\\nonumber&\times&m^7
\alpha+2 t^{3 m+7} \alpha \bigg(4
\bigg(\frac{m^2}{t^2}\bigg)^{\frac{\psi}{2}} t^{3 m} \alpha -\xi
(\psi -4)\bigg) (\psi -2) \psi
\bigg(\frac{m^2}{t^2}\bigg)^{\frac{\psi +1}{2}}
\\\nonumber&+&72 m^3 t^{3 m+4}\alpha  \psi  \bigg(2 t^{3 m} \alpha
\bigg(\frac{m^2}{t^2}\bigg)^{\psi /2}+\xi -\xi  \psi \bigg)
\bigg(\frac{m^2}{t^2}\bigg)^{\frac{\psi +1}{2}}+m^2 t^{3 m+4}
\\\nonumber&\times&\bigg(t^{3 m+1} \alpha  \bigg(2 \alpha  \beta
\bigg(4 \sqrt{\frac{m^2}{t^2}} \beta  \psi ^4+\bigg(8-4
\sqrt{\frac{m^2}{t^2}} \beta \bigg) \psi ^3-\bigg(4
\sqrt{\frac{m^2}{t^2}} \beta +13\bigg) \\\nonumber&\times& \psi
^2+\bigg(4 \sqrt{\frac{m^2}{t^2}} \beta -3\bigg)\psi +3\bigg)
\bigg(\frac{m^2}{t^2}\bigg)^{\frac{\psi}{2}}+\sqrt{6} (\psi -1)
\bigg(3+\sqrt{\frac{m^2}{t^2}} \beta  \psi ^2\\\nonumber&+&4
\sqrt{\frac{m^2}{t^2}} \beta  \psi +\psi \bigg) c\bigg)
\bigg(\frac{m^2}{t^2}\bigg)^{\frac{\psi}{2}}+24 \alpha \xi  (\psi
-2) \psi \bigg(\frac{m^2}{t^2}\bigg)^{\frac{\psi +1}{2}}-48 t^{3 m}
\alpha ^2 \\\nonumber&\times& \psi \bigg(\frac{m^2}{t^2}\bigg)^{\psi
+\frac{1}{2}}-t \xi (\psi -1) \bigg(2
\bigg(\frac{m^2}{t^2}\bigg)^{\frac{\psi}{2}} \alpha\beta \big(\psi
^3-3 \psi ^2-\psi +3\big)-3 \sqrt{6} c\bigg)\bigg)\\\nonumber&-&12
 \bigg(\bigg(\frac{m^2}{t^2}\bigg)^{\frac{\psi}{2}}
t^{3 m} \alpha \bigg(2 \alpha
\beta\bigg(\frac{m^2}{t^2}\bigg)^{\frac{\psi}{2}} \bigg(4 \psi ^2+2
\sqrt{\frac{m^2}{t^2}} \beta  \bigg(\psi ^2-1\bigg) \psi
-1\bigg)-c\\\nonumber&\times&(\psi -1) \bigg(\sqrt{\frac{m^2}{t^2}}
\beta \psi ^2+2 \sqrt{\frac{m^2}{t^2}} \beta  \psi +\psi +1\bigg)
\sqrt{6}\bigg)-\xi (\psi -1) \bigg(\big(\psi
^2-1\big)\\\nonumber&\times&2 \alpha \beta
\bigg(\frac{m^2}{t^2}\bigg)^{\psi /2}+\sqrt{6} c\bigg)\bigg)m^4 t^{3
m+2}+72 m^5 t^2 \bigg(-t^{3 m} (\psi -1)\alpha  \xi \bigg( (\psi
+1)\\\nonumber&\times&\beta \psi-8 \sqrt{\frac{m^2}{t^2}}\bigg)
\bigg(\frac{m^2}{t^2}\bigg)^{\psi /2}+t^{6 m} \alpha  \bigg(2
\bigg(\frac{m^2}{t^2}\bigg)^{\psi /2} \alpha
\bigg(\sqrt{\frac{m^2}{t^2}} \beta ^2 \psi ^3+2 \beta  \psi
^2\\\nonumber&-&\sqrt{\frac{m^2}{t^2}} \bigg(\beta ^2-2\bigg) \psi
-2 \sqrt{\frac{m^2}{t^2}}\bigg)-\sqrt{6} (\psi -1) \psi
\bigg(\sqrt{\frac{m^2}{t^2}} \beta  (\psi +1)+1\bigg)
c\bigg)\\\nonumber&\times& \bigg(\frac{m^2}{t^2}\bigg)^{\psi /2}+4
\xi ^2 (\psi -1)
\sqrt{\frac{m^2}{t^2}}\bigg)\bigg)^2\bigg\}\bigg\{165888 m^{12} t^2
(\psi -1)^2 \bigg(t^{3 m} \alpha \\\nonumber&\times&
\bigg(\sqrt{\frac{m^2}{t^2}} \beta \psi +1\bigg)
\bigg(\frac{m^2}{t^2}\bigg)^{\psi /2}+\xi \bigg)^4\bigg\}^{-1}+
\bigg\{t^{3 m+1} \bigg(72 m^3 t^4 \alpha  \psi \bigg(-t^{3 m} \alpha
\xi \\\nonumber&\times& \big(\psi ^2+7 \psi -14\big)
\bigg(\frac{m^2}{t^2}\bigg)^{\frac{\psi}{2}}+8 t^{6 m} \alpha ^2
\bigg(\frac{m^2}{t^2}\bigg)^{\psi }+\xi ^2 \big(\psi ^2-5 \psi
+6\big)\bigg) \bigg(\frac{m^2}{t^2}\bigg)^{\frac{\psi
+1}{2}}\\\nonumber&+&2 t^7 \alpha  (\psi -2) \psi  \bigg(-t^{3 m}
\alpha  \xi  \bigg(\psi ^2+10 \psi -48\bigg)
\bigg(\frac{m^2}{t^2}\bigg)^{\psi /2}+24 t^{6 m} \alpha ^2
\bigg(\frac{m^2}{t^2}\bigg)^{\psi }\\\nonumber&+&\xi ^2 \bigg(\psi
^2-10 \psi +24\bigg)\bigg) \bigg(\frac{m^2}{t^2}\bigg)^{\frac{\psi
+1}{2}}+72 m^7 t^{6 m} \alpha ^2 \beta ^2 (\psi -1) \psi ^2 \bigg(8
\bigg(\frac{m^2}{t^2}\bigg)^{\psi /2}\\\nonumber&\times& \alpha
\beta  (\psi +1)-\sqrt{6} (\psi +2)^2 c\bigg)
\bigg(\frac{m^2}{t^2}\bigg)^{\psi }-12 m^6 t^{6 m} \alpha ^2 \beta
^2 (\psi -1) \psi ^2 \bigg(16
\\\nonumber&\times&\bigg(\frac{m^2}{t^2}\bigg)^{\psi /2}   (\psi +1)-\sqrt{6} \bigg(\psi ^2+6 \psi +8\bigg) c\bigg)
\bigg(\frac{m^2}{t^2}\bigg)^{\psi }+m^2 t^4 \bigg(-t^{3 m+1} \alpha
\xi \\\nonumber&\times&\bigg(2 \alpha  \beta
\bigg(\sqrt{\frac{m^2}{t^2}} \beta \psi ^6+\bigg(11
\sqrt{\frac{m^2}{t^2}} \beta +2\bigg) \psi ^5+\bigg(19-50
\sqrt{\frac{m^2}{t^2}} \beta \bigg) \psi ^4+2
\\\nonumber&\times&\bigg(13 \sqrt{\frac{m^2}{t^2}} \beta -62\bigg)
\psi ^3+\bigg(49 \sqrt{\frac{m^2}{t^2}} \beta +137\bigg) \psi
^2+\bigg(38-37 \sqrt{\frac{m^2}{t^2}} \beta \bigg) \psi
\\\nonumber&-&30\bigg) \bigg(\frac{m^2}{t^2}\bigg)^{\psi
/2}+\sqrt{6} (\psi -1) \bigg(\sqrt{\frac{m^2}{t^2}} \beta  \psi
^3+\bigg(1-6 \sqrt{\frac{m^2}{t^2}} \beta \bigg) \psi ^2-\bigg(37
\\\nonumber&\times&\sqrt{\frac{m^2}{t^2}} \beta +8\bigg) \psi -30\bigg) c\bigg)
\bigg(\frac{m^2}{t^2}\bigg)^{\psi /2}-24 \alpha  \xi ^2 \psi
\bigg(\psi ^2-6 \psi +8\bigg)
\bigg(\frac{m^2}{t^2}\bigg)^{\frac{\psi +1}{2}}\\\nonumber&-&192
t^{6 m} \alpha ^3 \psi  \bigg(\frac{m^2}{t^2}\bigg)^{\frac{3 \psi
}{2}+\frac{1}{2}}+\bigg(2 \alpha  \beta \bigg(72
\sqrt{\frac{m^2}{t^2}} \beta  \psi ^4+\bigg(72-107
\sqrt{\frac{m^2}{t^2}} \beta \bigg) \psi ^3\\\nonumber&-&\bigg(37
\sqrt{\frac{m^2}{t^2}} \beta +129\bigg) \psi ^2+\bigg(37
\sqrt{\frac{m^2}{t^2}} \beta -15\bigg) \psi +15\bigg)
\bigg(\frac{m^2}{t^2}\bigg)^{\psi /2}+\sqrt{6}\\\nonumber&\times&
(\psi -1) \bigg(2 \sqrt{\frac{m^2}{t^2}} \beta  \psi ^3+\bigg(18
\sqrt{\frac{m^2}{t^2}} \beta +1\bigg) \psi ^2+\bigg(37
\sqrt{\frac{m^2}{t^2}} \beta +8\bigg) \psi \\\nonumber&+&15\bigg)
c\bigg)t^{6 m+1} \alpha ^2 \bigg(\frac{m^2}{t^2}\bigg)^{\psi }+24
t^{3 m} \alpha ^2 \xi  \psi \bigg(\psi ^2+6 \psi -16\bigg)
\bigg(\frac{m^2}{t^2}\bigg)^{\psi +\frac{1}{2}}
+t\\\nonumber&\times&\xi ^2 (\psi -1) \bigg(2 \alpha  \beta
\big(\psi ^4-8 \psi ^3+14 \psi ^2+8 \psi -15\big)
\bigg(\frac{m^2}{t^2}\bigg)^{\psi /2}+15 \sqrt{6}
c\bigg)\bigg)\\\nonumber&+&72 m^5 t^2 \bigg( \alpha  \xi
\bigg(\sqrt{6} (\psi -1) \bigg(\sqrt{\frac{m^2}{t^2}} \beta  \psi
^3+\psi ^2-\bigg(3 \sqrt{\frac{m^2}{t^2}} \beta +2\bigg) \psi
-2\bigg) c\\\nonumber&-&\bigg(\frac{m^2}{t^2}\bigg)^{\psi /2} \alpha
\beta \bigg(\sqrt{\frac{m^2}{t^2}} \beta  \psi ^5+\bigg(9
\sqrt{\frac{m^2}{t^2}} \beta +2\bigg) \psi ^4+\bigg(16-7
\sqrt{\frac{m^2}{t^2}} \beta \bigg) \psi ^3\\\nonumber&-&\bigg(9
\sqrt{\frac{m^2}{t^2}} \beta +23\bigg) \psi ^2+\bigg(6
\sqrt{\frac{m^2}{t^2}} \beta -3\bigg) \psi +4\bigg)\bigg)t^{3 m}
\bigg(\frac{m^2}{t^2}\bigg)^{\psi /2}+t^{6 m}
\\\nonumber&\times&\alpha ^2 \bigg(2 \bigg(\frac{m^2}{t^2}\bigg)^{\psi
/2} \alpha  \beta \bigg(12 \psi ^2+3 \sqrt{\frac{m^2}{t^2}} \beta
\bigg(4 \psi ^2-1\bigg) \psi -1\bigg)-\sqrt{6} (\psi
-1)\\\nonumber&\times&\bigg(2 \sqrt{\frac{m^2}{t^2}} \beta \psi
^3+\bigg(6 \sqrt{\frac{m^2}{t^2}} \beta +1\bigg) \psi ^2+\bigg(3
\sqrt{\frac{m^2}{t^2}} \beta +2\bigg) \psi +1\bigg) c\bigg)
\bigg(\frac{m^2}{t^2}\bigg)^{\psi }\\\nonumber&+&\xi ^2 (\psi -1)
\bigg(\bigg(\frac{m^2}{t^2}\bigg)^{\psi /2} \alpha  \beta \bigg(\psi
^3-2 \psi ^2-\psi +2\bigg)-\sqrt{6} c\bigg)\bigg)+m^4 t^2 \bigg(12
\\\nonumber&\times&  \xi  \bigg(2 \bigg(\frac{m^2}{t^2}\bigg)^{\psi /2}
\alpha  \beta \bigg(\sqrt{\frac{m^2}{t^2}} \beta  \psi ^5+\bigg(8
\sqrt{\frac{m^2}{t^2}} \beta +2\bigg) \psi ^4-2 \bigg(5
\sqrt{\frac{m^2}{t^2}} \beta -7\bigg) \psi ^3\\\nonumber&-&4 \bigg(2
\sqrt{\frac{m^2}{t^2}} \beta +7\bigg) \psi ^2+\bigg(9
\sqrt{\frac{m^2}{t^2}} \beta -4\bigg) \psi +6\bigg)-\sqrt{6} (\psi
-1) \bigg(\sqrt{\frac{m^2}{t^2}} \beta  \psi
^3\\\nonumber&+&\bigg(1-2 \sqrt{\frac{m^2}{t^2}} \beta \bigg) \psi
^2-\bigg(9 \sqrt{\frac{m^2}{t^2}} \beta +4\bigg) \psi -6\bigg)
c\bigg)t^{3 m} \alpha \bigg(\frac{m^2}{t^2}\bigg)^{\psi /2}+t^{6
m+1}\\\nonumber&\times& \alpha ^2 \beta ^2 (\psi -1) \psi ^2
\bigg(48 \alpha  \beta \bigg(\psi ^2-1\bigg)
\bigg(\frac{m^2}{t^2}\bigg)^{\psi /2}+\sqrt{6} \bigg(\psi ^2+10 \psi
+24\bigg) c\bigg)\\\nonumber&\times&
\bigg(\frac{m^2}{t^2}\bigg)^{\psi }-12 t^{6 m} \alpha ^2 \bigg(6
\bigg(\frac{m^2}{t^2}\bigg)^{\psi /2} \alpha \beta \bigg(8 \psi
^2+\sqrt{\frac{m^2}{t^2}} \beta  \bigg(8 \psi ^2-3\bigg) \psi
-1\bigg)\\\nonumber&-&\bigg(2 \sqrt{\frac{m^2}{t^2}} \beta \psi
^3+\bigg(10 \sqrt{\frac{m^2}{t^2}} \beta +1\bigg) \psi ^2+\bigg(9
\sqrt{\frac{m^2}{t^2}} \beta +4\bigg) \psi +3\bigg)(\psi
-1)\\\nonumber&\times& \sqrt{6}c\bigg)
\bigg(\frac{m^2}{t^2}\bigg)^{\psi }-12 \xi ^2  \bigg(2
\bigg(\frac{m^2}{t^2}\bigg)^{\psi /2} \alpha \beta \bigg(\psi ^3-3
\psi ^2-\psi +3\bigg)-3 \sqrt{6} c\bigg)\\\nonumber&\times&(\psi
-1)\bigg)\bigg)\bigg\}\bigg\{ m^8 \sqrt{\frac{m^2}{t^2}} (\psi -1)
\bigg(t^{3 m} \alpha \bigg(\sqrt{\frac{m^2}{t^2}} \beta \psi
+1\bigg) \bigg(\frac{m^2}{t^2}\bigg)^{\psi /2}+\xi \bigg)^3
\\\nonumber&\times&6912\bigg\}^{-1} +\bigg\{t^{3m}\bigg(72 m^5 \alpha  \psi \bigg(t^{3
m} \bigg(2 \bigg(\frac{m^2}{t^2}\bigg)^{\frac{\psi}{2}} \alpha \beta
\bigg(2 \psi +\sqrt{\frac{m^2}{t^2}} \beta  \big(\psi
^2-1\big)\bigg)\\\nonumber&-&\sqrt{6} (\psi -1)
\bigg(\sqrt{\frac{m^2}{t^2}} \beta  (\psi +1)+1\bigg) c\bigg)-\beta
\xi  \bigg(\psi ^2-1\bigg)\bigg) \bigg(\frac{m^2}{t^2}\bigg)^{\psi
/2}+2 t^5 \alpha \\\nonumber&\times&\bigg(4
\bigg(\frac{m^2}{t^2}\bigg)^{\psi /2} t^{3 m} \alpha -\xi (\psi
-4)\bigg) (\psi -2) \psi \bigg(\frac{m^2}{t^2}\bigg)^{\frac{\psi
+1}{2}}+72 m^3 t^2 \alpha \psi  \bigg(2 t^{3 m}\\\nonumber&\times&
\alpha \bigg(\frac{m^2}{t^2}\bigg)^{\psi /2}+\xi -\xi  \psi \bigg)
\bigg(\frac{m^2}{t^2}\bigg)^{\frac{\psi +1}{2}}+m^2 t^2 \bigg(t^{3
m+1} \alpha  \bigg(2 \alpha  \beta \bigg(4 \sqrt{\frac{m^2}{t^2}}
\beta  \psi ^4\\\nonumber&+&\bigg(8-4 \sqrt{\frac{m^2}{t^2}} \beta
\bigg) \psi ^3-\bigg(4 \sqrt{\frac{m^2}{t^2}} \beta +13\bigg) \psi
^2+\bigg(4 \sqrt{\frac{m^2}{t^2}} \beta -3\bigg) \psi
+3\bigg)\\\nonumber&\times& \bigg(\frac{m^2}{t^2}\bigg)^{\psi
/2}+\sqrt{6} (\psi -1) \bigg(\sqrt{\frac{m^2}{t^2}} \beta  \psi ^2+4
\sqrt{\frac{m^2}{t^2}} \beta  \psi +\psi +3\bigg) c\bigg)
\bigg(\frac{m^2}{t^2}\bigg)^{\psi /2}\\\nonumber&+&24 \alpha  \xi
(\psi -2) \psi \bigg(\frac{m^2}{t^2}\bigg)^{\frac{\psi +1}{2}}-48
t^{3 m} \alpha ^2 \psi  \bigg(\frac{m^2}{t^2}\bigg)^{\psi
+\frac{1}{2}}-t \xi (\psi -1) \bigg(2 \alpha \beta
\\\nonumber&\times& \bigg(\frac{m^2}{t^2}\bigg)^{\psi /2} \alpha
\beta \bigg(\psi ^3-3 \psi ^2-\psi +3\bigg)-3 \sqrt{6}
c\bigg)\bigg)-12 m^4 \bigg(\bigg(\frac{m^2}{t^2}\bigg)^{\psi /2}
\\\nonumber&\times&t^{3 m} \alpha \bigg( \bigg(\frac{m^2}{t^2}\bigg)^{\psi /2} 2\alpha
\beta \bigg(4 \psi ^2+2 \sqrt{\frac{m^2}{t^2}} \beta \big(\psi
^2-1\big) \psi -1\bigg)-\sqrt{6} (\psi -1)\\\nonumber&\times&
\bigg(\sqrt{\frac{m^2}{t^2}} \beta \psi ^2+2 \sqrt{\frac{m^2}{t^2}}
\beta  \psi +\psi +1\bigg) c\bigg)-\xi  \bigg(\sqrt{6} c+ \big(\psi
^2-1\big) \bigg(\frac{m^2}{t^2}\bigg)^{\psi /2}\\\nonumber&\times&2
\alpha \beta \bigg)(\psi -1)\bigg)\bigg)\bigg\}\bigg\{576
\sqrt{\frac{m^2}{t^2}}  \bigg(t^{3 m} \alpha
\bigg(\sqrt{\frac{m^2}{t^2}} \beta \psi +1\bigg)
\bigg(\frac{m^2}{t^2}\bigg)^{\psi /2}+\xi \bigg)^2
\\\nonumber&\times& (\psi -1) m^5\bigg\}+\frac{1}{2} ,
\\\nonumber s&=&\bigg\{192 m
\bigg(\frac{m^2}{t^2}\bigg)^{5/2} t^{4-3 m} (\psi -1) \bigg(t^{3 m}
\alpha  \bigg(\sqrt{\frac{m^2}{t^2}} \beta  \psi +1\bigg)
\bigg(\frac{m^2}{t^2}\bigg)^{\psi /2}+\xi \bigg)^2
\\\nonumber&\times&\bigg(\bigg\{\bigg(288 m^7   \beta  (\psi -1) \psi
\bigg(t^{3 m} \alpha  \bigg(\sqrt{\frac{m^2}{t^2}} \beta  \psi
+2\bigg) \bigg(\frac{m^2}{t^2}\bigg)^{\psi /2}+2 \xi \bigg)t^{3 m}
\alpha \\\nonumber&\times&\bigg(\frac{m^2}{t^2}\bigg)^{\psi /2}+2
t^{3 m+7} \alpha \bigg(4 \bigg(\frac{m^2}{t^2}\bigg)^{\psi /2} t^{3
m} \alpha -\xi (\psi -4)\bigg) (\psi -2) \psi
\bigg(\frac{m^2}{t^2}\bigg)^{\frac{\psi +1}{2}}\\\nonumber&+&72 m^3
t^{3 m+4} \alpha  \psi  \bigg(2 t^{3 m} \alpha
\bigg(\frac{m^2}{t^2}\bigg)^{\psi /2}+\xi -\xi  \psi \bigg)
\bigg(\frac{m^2}{t^2}\bigg)^{\frac{\psi +1}{2}}+m^2 t^{3 m+4}
\\\nonumber&\times&\bigg(t^{3 m+1}   \bigg(2 \alpha  \beta  \bigg(4
\sqrt{\frac{m^2}{t^2}} \beta  \psi ^4+\bigg(8-4
\sqrt{\frac{m^2}{t^2}} \beta \bigg) \psi ^3-\bigg(4
\sqrt{\frac{m^2}{t^2}} \beta +13\bigg) \psi ^2\\\nonumber&+&\bigg(4
\sqrt{\frac{m^2}{t^2}} \beta -3\bigg) \psi +3\bigg)
\bigg(\frac{m^2}{t^2}\bigg)^{\psi /2}+\sqrt{6} (\psi -1) \bigg(\psi
+3+\sqrt{\frac{m^2}{t^2}} \beta  \psi ^2\\\nonumber&+&4
\sqrt{\frac{m^2}{t^2}} \beta  \psi \bigg) c\bigg)
\alpha\bigg(\frac{m^2}{t^2}\bigg)^{\psi /2}+24 \alpha  \xi  (\psi
-2) \psi \bigg(\frac{m^2}{t^2}\bigg)^{\frac{\psi +1}{2}}-48
\bigg(\frac{m^2}{t^2}\bigg)^{\psi +\frac{1}{2}}\\\nonumber&\times&
t^{3 m}\alpha ^2\psi-t \xi (\psi -1) \bigg(2
\bigg(\frac{m^2}{t^2}\bigg)^{\psi /2} \alpha \beta \bigg(\psi ^3-3
\psi ^2-\psi +3\bigg)-3 \sqrt{6} c\bigg)\bigg)\\\nonumber&-&12 m^4
t^{3 m+2} \bigg(\bigg(\frac{m^2}{t^2}\bigg)^{\frac{\psi}{2}} t^{3 m}
\alpha \bigg(2 \bigg(\frac{m^2}{t^2}\bigg)^{\frac{\psi}{2}}\bigg(4
\psi ^2+2 \sqrt{\frac{m^2}{t^2}} \beta  \big(\psi ^2-1\big) \psi
-1\bigg)\\\nonumber&\times&\alpha \beta-\sqrt{6} (\psi -1)
\bigg(\sqrt{\frac{m^2}{t^2}} \beta \psi ^2+2 \sqrt{\frac{m^2}{t^2}}
\beta  \psi +\psi +1\bigg) c\bigg)-\xi (\psi -1)
\\\nonumber&\times&\bigg(2 \alpha \beta  \big(\psi ^2-1\big)
\bigg(\frac{m^2}{t^2}\bigg)^{\psi /2}+\sqrt{6} c\bigg)\bigg)+72 m^5
t^2 \bigg(-t^{3 m} \alpha \bigg(\beta  \psi (\psi +1)\\\nonumber&-&8
\sqrt{\frac{m^2}{t^2}}\bigg) \xi (\psi
-1)\bigg(\frac{m^2}{t^2}\bigg)^{\psi /2}+t^{6 m} \alpha  \bigg(2
\bigg(\frac{m^2}{t^2}\bigg)^{\psi /2} \alpha
\bigg(\sqrt{\frac{m^2}{t^2}} \beta ^2 \psi ^3+2 \beta  \psi
^2\\\nonumber&-&\sqrt{\frac{m^2}{t^2}} \big(\beta ^2-2\big) \psi -2
\sqrt{\frac{m^2}{t^2}}\bigg)-\sqrt{6} (\psi -1) \psi
\bigg(\sqrt{\frac{m^2}{t^2}} \beta  (\psi +1)+1\bigg) c\bigg)
\\\nonumber&\times&\bigg(\frac{m^2}{t^2}\bigg)^{\psi /2}+4 \xi ^2 (\psi -1)
\sqrt{\frac{m^2}{t^2}}\bigg)\bigg)^2\bigg\}\bigg\{165888 (\psi -1)^2
\bigg(\bigg(\sqrt{\frac{m^2}{t^2}} \beta \psi
+1\bigg)\\\nonumber&\times&t^{3 m} \alpha
\bigg(\frac{m^2}{t^2}\bigg)^{\psi /2}+\xi \bigg)^4m^{12}
t^2\bigg\}^{-1}+\bigg\{t^{3 m+1} \bigg(72 m^3 t^4 \alpha  \psi
\bigg(-t^{3 m}\bigg(\frac{m^2}{t^2}\bigg)^{\psi
/2}\\\nonumber&\times&\alpha \xi \big(\psi ^2+7 \psi -14\big) +8
t^{6 m} \alpha ^2 \bigg(\frac{m^2}{t^2}\bigg)^{\psi }+\xi ^2
\bigg(\psi ^2-5 \psi +6\bigg)\bigg)
\bigg(\frac{m^2}{t^2}\bigg)^{\frac{\psi +1}{2}}\\\nonumber&+&2 t^7
\alpha  (\psi -2) \psi  \bigg(-t^{3 m} \alpha  \xi \bigg(\psi ^2+10
\psi -48\bigg) \bigg(\frac{m^2}{t^2}\bigg)^{\psi /2}+24 t^{6 m}
\alpha ^2 \bigg(\frac{m^2}{t^2}\bigg)^{\psi }\\\nonumber&+&\xi ^2
\bigg(\psi ^2-10 \psi +24\bigg)\bigg)
\bigg(\frac{m^2}{t^2}\bigg)^{\frac{\psi +1}{2}}+72 m^7 t^{6 m}
\alpha ^2  (\psi -1) \psi ^2 \bigg(8
\bigg(\frac{m^2}{t^2}\bigg)^{\psi /2} \\\nonumber&\times&\alpha
\beta  (\psi +1)-\sqrt{6} (\psi +2)^2 c\bigg) \beta
^2\bigg(\frac{m^2}{t^2}\bigg)^{\psi }-12 m^6 t^{6 m} \alpha ^2 \beta
^2 (\psi -1) \psi ^2 \\\nonumber&\times&\bigg(16
\bigg(\frac{m^2}{t^2}\bigg)^{\psi /2} \alpha  \beta  (\psi
+1)-\sqrt{6} \bigg(\psi ^2+6 \psi +8\bigg) c\bigg)
\bigg(\frac{m^2}{t^2}\bigg)^{\psi }+m^2 t^4
\\\nonumber&\times&\bigg(\bigg(11
\sqrt{\frac{m^2}{t^2}} \beta +2\bigg) \psi ^5- \xi \bigg(2 \alpha
\beta \bigg(\sqrt{\frac{m^2}{t^2}} \beta \psi ^6+\bigg(19-50
\sqrt{\frac{m^2}{t^2}} \beta \bigg) \psi ^4\\\nonumber&+&2 \bigg(13
\sqrt{\frac{m^2}{t^2}} \beta -62\bigg) \psi ^3+\bigg(49
\sqrt{\frac{m^2}{t^2}} \beta +137\bigg) \psi ^2+\bigg(38-37
\sqrt{\frac{m^2}{t^2}} \beta \bigg) \psi \\\nonumber&-&30\bigg)
\bigg(\frac{m^2}{t^2}\bigg)^{\psi /2}+\sqrt{6} (\psi -1)
\bigg(\sqrt{\frac{m^2}{t^2}} \beta  \psi ^3+\bigg(1-6
\sqrt{\frac{m^2}{t^2}} \beta \bigg) \psi ^2-30\\\nonumber&-&\bigg(37
\sqrt{\frac{m^2}{t^2}} \beta +8\bigg) \psi \bigg) c\bigg)t^{3 m+1}
\alpha \bigg(\frac{m^2}{t^2}\bigg)^{\psi /2}- \psi\big(\psi ^2-6
\psi +8\big) \bigg(\frac{m^2}{t^2}\bigg)^{\frac{\psi
+1}{2}}\\\nonumber&\times& 24 \alpha \xi ^2-192 t^{6 m} \alpha ^3
\psi \bigg(\frac{m^2}{t^2}\bigg)^{\frac{3 \psi }{2}+\frac{1}{2}}+
\alpha ^2 \bigg(2 \alpha  \beta \bigg(72 \sqrt{\frac{m^2}{t^2}}
\beta  \psi ^4+\bigg(72-107
\\\nonumber&\times&\sqrt{\frac{m^2}{t^2}} \beta \bigg) \psi
^3-\bigg(37 \sqrt{\frac{m^2}{t^2}} \beta +129\bigg) \psi ^2+\bigg(37
\sqrt{\frac{m^2}{t^2}} \beta -15\bigg) \psi +15\bigg)
\\\nonumber&\times&\bigg(\frac{m^2}{t^2}\bigg)^{\psi /2}+\sqrt{6} (\psi -1) \bigg(2
\sqrt{\frac{m^2}{t^2}} \beta  \psi ^3+\bigg(18
\sqrt{\frac{m^2}{t^2}} \beta +1\bigg) \psi
^2+15\\\nonumber&+&\bigg(37 \sqrt{\frac{m^2}{t^2}} \beta +8\bigg)
\psi \bigg) c\bigg)t^{6 m+1} \bigg(\frac{m^2}{t^2}\bigg)^{\psi }+24
 \big(\psi ^2+6 \psi -16\big)
\bigg(\frac{m^2}{t^2}\bigg)^{\psi
+\frac{1}{2}}\\\nonumber&\times&t^{3 m} \alpha ^2 \xi \psi+t \xi ^2
(\psi -1) \bigg(2 \alpha  \beta  \bigg(\psi ^4-8 \psi ^3+14 \psi
^2+8 \psi -15\bigg) \bigg(\frac{m^2}{t^2}\bigg)^{\psi
/2}\\\nonumber&+&15 \sqrt{6} c\bigg)\bigg)+72  \bigg(t^{3 m}
\bigg(\sqrt{6} (\psi -1) \bigg(\sqrt{\frac{m^2}{t^2}} \beta \psi
^3+\psi ^2-\bigg(3 \sqrt{\frac{m^2}{t^2}} \beta +2\bigg) \psi
\\\nonumber&-&2\bigg) c- \alpha  \beta
\bigg(\sqrt{\frac{m^2}{t^2}} \beta  \psi ^5+\bigg(9
\sqrt{\frac{m^2}{t^2}} \beta +2\bigg) \psi ^4+\bigg(16-7
\sqrt{\frac{m^2}{t^2}} \beta \bigg) \psi ^3\\\nonumber&-&\bigg(9
\sqrt{\frac{m^2}{t^2}} \beta +23\bigg) \psi ^2+\bigg(6
\sqrt{\frac{m^2}{t^2}} \beta -3\bigg) \psi
+4\bigg)\bigg(\frac{m^2}{t^2}\bigg)^{\psi /2}\bigg)\alpha \xi
\bigg(\frac{m^2}{t^2}\bigg)^{\psi /2}\\\nonumber&+&t^{6 m} \alpha ^2
\bigg(2 \bigg(\frac{m^2}{t^2}\bigg)^{\psi /2} \alpha  \beta \bigg(12
\psi ^2+3 \sqrt{\frac{m^2}{t^2}} \beta  \bigg(4 \psi ^2-1\bigg) \psi
-1\bigg)- (\psi -1) \\\nonumber&\times&\sqrt{6}\bigg(2
\sqrt{\frac{m^2}{t^2}} \beta \psi ^3+\bigg(6 \sqrt{\frac{m^2}{t^2}}
\beta +1\bigg) \psi ^2+\bigg(3 \sqrt{\frac{m^2}{t^2}} \beta +2\bigg)
\psi +1\bigg) c\bigg)\\\nonumber&\times&
\bigg(\frac{m^2}{t^2}\bigg)^{\psi }+\xi ^2 (\psi -1)
\bigg(\bigg(\frac{m^2}{t^2}\bigg)^{\frac{\psi}{2}} \alpha  \beta
\big(\psi ^3-2 \psi ^2-\psi +2\big)-\sqrt{6} c\bigg)\bigg)m^5
t^2\\\nonumber&+&m^4 t^2 \bigg(12 t^{3 m} \alpha  \xi  \bigg(2
\bigg(\frac{m^2}{t^2}\bigg)^{\psi /2} \alpha  \beta
\bigg(\sqrt{\frac{m^2}{t^2}} \beta  \psi ^5+\bigg(8
\sqrt{\frac{m^2}{t^2}} \beta +2\bigg) \psi ^4-2 \psi
^3\\\nonumber&\times& \bigg(5 \sqrt{\frac{m^2}{t^2}} \beta -7\bigg)
-4 \bigg(2 \sqrt{\frac{m^2}{t^2}} \beta +7\bigg) \psi ^2+\bigg(9
\sqrt{\frac{m^2}{t^2}} \beta -4\bigg) \psi
+6\bigg)\\\nonumber&-&\sqrt{6} \bigg(\sqrt{\frac{m^2}{t^2}} \beta
\psi ^3+\bigg(1-2 \sqrt{\frac{m^2}{t^2}} \beta \bigg) \psi
^2-\bigg(9 \sqrt{\frac{m^2}{t^2}} \beta +4\bigg) \psi \bigg)(\psi
-1) c\bigg) \\\nonumber&\times&\bigg(\frac{m^2}{t^2}\bigg)^{\psi
/2}+ \beta ^2  \bigg(48 \alpha \beta  \bigg(\psi ^2-1\bigg)
\bigg(\frac{m^2}{t^2}\bigg)^{\psi /2}+\sqrt{6} \bigg(\psi ^2+10 \psi
+24\bigg) c\bigg)\\\nonumber&\times&(\psi -1) \psi ^2 t^{6 m+1}
\alpha ^2 \bigg(\frac{m^2}{t^2}\bigg)^{\psi }-12 t^{6 m} \alpha ^2
\bigg(6 \bigg(\frac{m^2}{t^2}\bigg)^{\psi /2} \alpha \beta \bigg(8
\psi ^2+\sqrt{\frac{m^2}{t^2}} \beta \\\nonumber&\times& \big(8 \psi
^2-3\big) \psi -1\bigg)-\sqrt{6} (\psi -1) \bigg(2
\sqrt{\frac{m^2}{t^2}} \beta \psi ^3+\bigg(10 \sqrt{\frac{m^2}{t^2}}
\beta +1\bigg) \psi ^2\\\nonumber&+&\bigg(9 \sqrt{\frac{m^2}{t^2}}
\beta +4\bigg) \psi +3\bigg) c\bigg)
\bigg(\frac{m^2}{t^2}\bigg)^{\psi }-12 \xi ^2 (\psi -1) \bigg(2
\bigg(\frac{m^2}{t^2}\bigg)^{\psi /2} \alpha \beta
\\\nonumber&\times&\big(\psi ^3-3 \psi ^2-\psi +3\big)-3 \sqrt{6}
c\bigg)\bigg)\bigg)\bigg\}\bigg\{6912 (\psi -1) \bigg( \alpha
\bigg(\sqrt{\frac{m^2}{t^2}} \beta \psi
+1\bigg)\\\nonumber&\times&t^{3 m} \bigg(\frac{m^2}{t^2}\bigg)^{\psi
/2}+\xi \bigg)^3m^8 \sqrt{\frac{m^2}{t^2}}\bigg\}^{-1}+\bigg\{t^{3
m} \bigg(72 m^5 \alpha \psi \bigg(t^{3 m} \bigg(2
\bigg(\frac{m^2}{t^2}\bigg)^{\psi /2} \alpha\\\nonumber&\times&
\beta \bigg(2 \psi +\sqrt{\frac{m^2}{t^2}} \beta  \bigg(\psi
^2-1\bigg)\bigg)-\sqrt{6} (\psi -1) \bigg(\sqrt{\frac{m^2}{t^2}}
\beta  (\psi +1)+1\bigg) c\bigg)\\\nonumber&-&\beta  \xi \bigg(\psi
^2-1\bigg)\bigg) \bigg(\frac{m^2}{t^2}\bigg)^{\psi /2}+2 t^5 \alpha
\bigg(4 \bigg(\frac{m^2}{t^2}\bigg)^{\psi /2} t^{3 m} \alpha -\xi
(\psi -4)\bigg) (\psi -2)\\\nonumber&\times& \psi
\bigg(\frac{m^2}{t^2}\bigg)^{\frac{\psi +1}{2}}+72 m^3 t^2 \alpha
\psi  \bigg(2 t^{3 m} \alpha \bigg(\frac{m^2}{t^2}\bigg)^{\psi
/2}+\xi -\xi  \psi \bigg) \bigg(\frac{m^2}{t^2}\bigg)^{\frac{\psi
+1}{2}}+m^2 \\\nonumber&\times&t^2 \bigg( \alpha  \bigg(2 \alpha
\beta \bigg(4 \sqrt{\frac{m^2}{t^2}} \beta  \psi ^4+\bigg(8-4
\sqrt{\frac{m^2}{t^2}} \beta \bigg) \psi ^3-\bigg(4
\sqrt{\frac{m^2}{t^2}} \beta +13\bigg) \psi ^2\\\nonumber&+&\bigg(4
\sqrt{\frac{m^2}{t^2}} \beta -3\bigg) \psi +3\bigg)
\bigg(\frac{m^2}{t^2}\bigg)^{\psi /2}+ (\psi -1)
\bigg(\sqrt{\frac{m^2}{t^2}} \beta  \psi ^2+4 \sqrt{\frac{m^2}{t^2}}
\beta  \psi \\\nonumber&+&\psi +3\bigg) \sqrt{6} c\bigg)t^{3 m+1}
\bigg(\frac{m^2}{t^2}\bigg)^{\psi /2}+24 \alpha  \xi  (\psi -2) \psi
\bigg(\frac{m^2}{t^2}\bigg)^{\frac{\psi +1}{2}}-48 t^{3 m} \alpha ^2
\psi \\\nonumber&\times&\bigg(\frac{m^2}{t^2}\bigg)^{\psi
+\frac{1}{2}}-t \xi (\psi -1) \bigg(2
\bigg(\frac{m^2}{t^2}\bigg)^{\frac{\psi}{2}} \alpha \beta \big(\psi
^3-3 \psi ^2-\psi +3\big)-3 \sqrt{6} c\bigg)\bigg)\\\nonumber&-&12
m^4 \bigg(\bigg(\frac{m^2}{t^2}\bigg)^{\frac{\psi}{2}} t^{3 m}
\alpha \bigg(2 \bigg(\frac{m^2}{t^2}\bigg)^{\frac{\psi}{2}} \alpha
\beta \bigg(4 \psi ^2+2 \sqrt{\frac{m^2}{t^2}} \beta \bigg(\psi
^2-1\bigg) \psi -1\bigg)\\\nonumber&-&\sqrt{6} (\psi -1)
\bigg(\sqrt{\frac{m^2}{t^2}} \beta \psi ^2+2 \sqrt{\frac{m^2}{t^2}}
\beta  \psi +\psi +1\bigg) c\bigg)-\xi (\psi -1) \bigg(\big(\psi
^2-1\big)\\\nonumber&\times& 2 \alpha \beta
\bigg(\frac{m^2}{t^2}\bigg)^{\psi /2}+\sqrt{6}
c\bigg)\bigg)\bigg)\bigg\}\bigg\{ m^5 \sqrt{\frac{m^2}{t^2}} (\psi
-1) \bigg(t^{3 m} \alpha \bigg(\sqrt{\frac{m^2}{t^2}} \beta \psi
+1\bigg)\\\nonumber&\times& \bigg(\frac{m^2}{t^2}\bigg)^{\psi
/2}+\xi \bigg)^2 576\bigg\}^{-1}-\frac{1}{2}\bigg)\bigg\}\bigg\{72
m^5 \alpha \psi \bigg(t^{3 m} \bigg(2
\bigg(\frac{m^2}{t^2}\bigg)^{\psi /2} \alpha \beta \bigg(2 \psi
\\\nonumber&+&\sqrt{\frac{m^2}{t^2}} \beta  \big(\psi ^2-1\big)\bigg)-\sqrt{6}
(\psi -1) \bigg(\sqrt{\frac{m^2}{t^2}} \beta  (\psi +1)+1\bigg)
c\bigg)-\beta  \xi  \big(\psi ^2-1\big)\bigg)\\\nonumber&\times&
\bigg(\frac{m^2}{t^2}\bigg)^{\frac{\psi}{2}}+2 t^5 \alpha \bigg(4
\bigg(\frac{m^2}{t^2}\bigg)^{\frac{\psi}{2}} t^{3 m} \alpha -\xi
(\psi -4)\bigg) (\psi -2) \psi
\bigg(\frac{m^2}{t^2}\bigg)^{\frac{\psi +1}{2}}+
\alpha\\\nonumber&\times& \psi 72 m^3 t^2 \bigg(2 t^{3 m} \alpha
\bigg(\frac{m^2}{t^2}\bigg)^{\psi /2}+\xi -\xi  \psi \bigg)
\bigg(\frac{m^2}{t^2}\bigg)^{\frac{\psi +1}{2}}+m^2 t^2 \bigg(t^{3
m+1} \alpha  \bigg(2 \alpha  \beta \\\nonumber&\times&\bigg(4
\sqrt{\frac{m^2}{t^2}} \beta  \psi ^4+\bigg(8-4
\sqrt{\frac{m^2}{t^2}} \beta \bigg) \psi ^3-\bigg(4
\sqrt{\frac{m^2}{t^2}} \beta +13\bigg) \psi
^2+\bigg(3-3+4\\\nonumber&\times& \psi\sqrt{\frac{m^2}{t^2}} \beta
\bigg) \bigg) \bigg(\frac{m^2}{t^2}\bigg)^{\frac{\psi}{2}}+\sqrt{6}
\bigg(\sqrt{\frac{m^2}{t^2}} \beta  \psi ^2+4 \sqrt{\frac{m^2}{t^2}}
\beta  \psi +\psi +3\bigg)c(\psi -1)\bigg)\\\nonumber&\times&
\bigg(\frac{m^2}{t^2}\bigg)^{\frac{\psi}{2}}+24 \alpha  \xi  (\psi
-2) \psi \bigg(\frac{m^2}{t^2}\bigg)^{\frac{\psi +1}{2}}-48 t^{3 m}
\alpha ^2 \psi  \bigg(\frac{m^2}{t^2}\bigg)^{\psi +\frac{1}{2}}-
(\psi -1)t\\\nonumber&\times&  \xi\bigg(2
\bigg(\frac{m^2}{t^2}\bigg)^{\frac{\psi}{2}} \alpha \beta \big(\psi
^3-3 \psi ^2-\psi +3\big)-3 \sqrt{6} c\bigg)\bigg)-12
m^4\bigg(\bigg(\frac{m^2}{t^2}\bigg)^{\frac{\psi}{2}} t^{3 m} \alpha
\\\nonumber&\times& \bigg(2 \bigg(\frac{m^2}{t^2}\bigg)^{\frac{\psi}{2}}
\alpha \beta \bigg(4 \psi ^2+2 \sqrt{\frac{m^2}{t^2}} \beta-\sqrt{6}
(\psi -1) \bigg(\sqrt{\frac{m^2}{t^2}} \beta \psi ^2 \big(\psi
^2-1\big) \psi -1\bigg)\\\nonumber&+&2 \sqrt{\frac{m^2}{t^2}} \beta
\psi +\psi +1\bigg) c\bigg)-\xi (\psi -1) \bigg(2 \alpha \beta
\big(\psi ^2-1\big)
\bigg(\frac{m^2}{t^2}\bigg)^{\frac{\psi}{2}}+\sqrt{6}
c\bigg)\bigg)\bigg\}^{-1}.
\end{eqnarray}\\
\textbf{Data Availability Statement:} No data was used for the
research described in this paper.

\end{document}